\documentclass[10pt]{iopart}

\usepackage{amssymb}
\usepackage{psfrag}
\usepackage{harvard}
\usepackage{amstext}
\usepackage{epsfig}
\usepackage{rotating}
\usepackage{longtable}
\usepackage{multirow}
\usepackage{tabularx}
\usepackage[verbose]{layout}
\usepackage{afterpage}

\begin{document}

\title{Electromagnetically Induced Transparency in $^{6}$Li}

\author{J Fuchs, G J Duffy, W J Rowlands and A M Akulshin}

 \address{ARC Centre of Excellence for Quantum Atom Optics, Centre for Atom Optics and Ultrafast Spectroscopy, Swinburne
University of Technology, Melbourne, Victoria 3122, Australia}

\ead{jfuchs@swin.edu.au}

\begin{abstract}
We report electromagnetically induced transparency for the D1 and
D2 lines in $^{6}$Li in both a vapour cell and an atomic beam.
Electromagnetically induced transparency is created using
co-propagating mutually coherent laser beams with a frequency
difference equal to the hyperfine ground state splitting of
228.2~MHz. The effects of various optical polarization
configurations and applied magnetic fields are investigated. In
addition, we apply an optical Ramsey spectroscopy technique which
further reduces the observed resonance width.

\end{abstract}

 \submitto{\JPB} \maketitle

\maketitle

\section{Introduction}

Electromagnetically Induced Transparency (EIT) is of great
interest due to its wide application in lasing without inversion
\cite{Zibrov95}, control of light propagation (slow light)
\cite{Matsko01} and enhanced Kerr nonlinearity
\cite{Harris99,Akulshin03}. The concept of a non-absorbing dark
state is the key basis of light storage \cite{Lukin03}. Lithium
has two stable isotopes, $^{6}$Li and $^{7}$Li (7.5~$\%$ and
92.5~$\%$ natural abundance, respectively). Magnus {\emph et~al}
\cite{Magnus05} reported the first study of EIT on the D1 and D2
lines of $^{7}$Li in a vapour. In this paper we present EIT in
$^{6}$Li performed on both the D1 and D2 lines using several
nonlinear techniques, paying particular attention to a comparison
between the EIT resonances obtained on the D1 and D2 lines.

EIT is a phenomenon in which an opaque atomic medium is turned
into a transparent one in the presence of a control laser field
\cite{Harris97}. When two Zeeman sub-levels or two hyperfine
levels in an atomic ground state are coupled by light to a common
excited state, the interference between amplitudes of alternative
transition paths can substantially reduce the absorption. The
atoms are pumped by the laser light into a coherent superposition
of the ground state sublevels which constitutes a non-absorbing
state decoupled from the laser field. This phenomenon is also
known as coherent population trapping, e.g.,\cite{Arimondo96}. The
absorption of the probe exhibits a narrow dip that has a
sub-natural linewidth ultimately determined by the relaxation time
of the ground state sublevels. The refractive index of the
coherent atomic medium reveals extremely steep normal dispersion
in the vicinity of the EIT resonance, so that the group velocity
of light in such an atomic medium can be dramatically reduced
\cite{Schmidt96}.

\begin{figure}[htb]
\begin{center}
\psfrag{D1=670.979 nm}{\raisebox{0mm}{\small{D1 670.979~nm}}}
\psfrag{D2=670.977 nm}{\raisebox{0mm}{\small{D2 670.977~nm}}}

\psfrag{26.1 MHz}{\raisebox{0mm}{\small{26.1~MHz}}} \psfrag{10.053
GHz}{\raisebox{0mm}{\small{10.053~GHz}}} \psfrag{228.2
MHz}{\raisebox{0mm}{\small{228.2~MHz}}}

\psfrag{4.6 MHz}{\raisebox{0mm}{\small{4.6~MHz}}}

\psfrag{10.056 GHz}{\raisebox{0mm}{\small{10.056~GHz}}}
\psfrag{F=1/2}{\raisebox{0mm}{\hspace{3mm}\small{F=1/2}}}
\psfrag{F=3/2}{\raisebox{0mm}{\hspace{3mm}\small{F=3/2}}}
\psfrag{F'=1/2}{\raisebox{0mm}{\hspace{3mm}\small{F$'$=1/2}}}
\psfrag{F'=3/2}{\raisebox{0mm}{\hspace{3mm}\small{F$'$=3/2}}}
\psfrag{F'=5/2}{\raisebox{0mm}{\hspace{3mm}\small{F$'$=5/2}}}
\psfrag{22P3/2}{\raisebox{1mm}{\hspace{3mm}\small{2$^{2}P_{3/2}$}}}
\psfrag{22P1/2}{\raisebox{0mm}{\hspace{3mm}\small{2$^{2}P_{1/2}$}}}
\psfrag{22S1/2}{\raisebox{0mm}{\hspace{3mm}\small{2$^{2}S_{1/2}$}}}
\epsfig{file=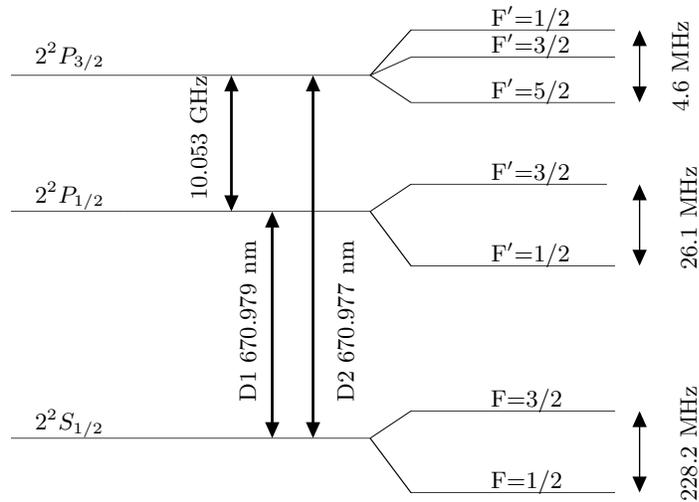,width=0.7\linewidth,height=0.5\linewidth}
\caption{A schematic energy level diagram for $^{6}$Li.}
\label{fig:energy_level}
\end{center}
\end{figure}

EIT in alkali atoms with half-integer total angular momentum
(F=J+I, J=L$\pm$1/2) have not, to our knowledge, been previously
investigated. A schematic diagram of the relevant atomic energy
levels is shown in figure~\ref{fig:energy_level}. The level
structure of $^{6}$Li is somewhat different to other alkali atoms
that have been used to study EIT and other effects related to
ground state coherence. There are only three sublevels in the
2$^{2}P_{3/2}$ state in comparison with four sublevels for all
other alkali n$^{2}P_{3/2}$ states.

The strongest optical transitions of $^{6}$Li are two electric
dipole transitions 2$^{2}S_{1/2}\rightarrow$ 2$^{2}P_{1/2}$ (D1
line) and 2$^{2}S_{1/2}\rightarrow$ 2$^{2}P_{3/2}$ (D2 line) which
are separated by a fine structure splitting of $\sim$ 10~GHz. Of
all the alkali atoms $^6$Li has the smallest hyperfine splitting
in both the ground and excited states. The hyperfine splitting of
the 2$^{2}S_{1/2}$ ground state, 228.2~MHz \cite{Walls03}, is much
less than the Doppler width of the D lines near room temperature.
But, more interestingly, the hyperfine splitting of the
2$^{2}P_{3/2}$ state is smaller than the 5.9~MHz
\cite{McAlexander96} natural width of the optical transition.
Thus, every atom has comparable probabilities, independent of the
atomic velocity, of being excited via different transitions on the
D2 line. This unique situation for alkali atoms allows analyzing a
role of additional optical transitions, which do not contribute to
the preparation of a non-absorbing coherent state responsible for
EIT. This study could be useful for optimization of EIT resonances
widely used for metrological applications \cite{Knappe05}.

The ultimate width of an EIT resonance also depends on the mutual
coherence of the probe and control laser fields \cite{Arimondo96}.
In the case of Cs and Rb, which have large ground state splitting,
several methods can be used to produce two phase-stable laser
fields such as high frequency acousto-optical modulators (AOM),
two phase-locked lasers, applying current modulation to laser
diodes and the use of electro-optic modulators \cite{Wynands99}.
High frequency AOMs are expensive while phase locking at precise
frequency offset and current modulation at high frequencies are
experimentally not simple. However, in the case of lithium the
ground state splitting is relatively small and mutually coherent
probe and control fields can be easily prepared from the same
laser using a low-cost AOM in a double pass configuration, thus
eliminating laser linewidth contribution.

\section{Experimental Set-up}

An external cavity diode laser (ECDL) (Toptica DL 100) is used as
the source of the resonant optical field which has an output power
of 15~mW and a linewidth of approximately 1~MHz. Using the
so-called feed-forward technique, where both the diode current and
the grating position are varied, a fine frequency tuning range of
over 20~GHz is obtained. The vapour cell consists of a stainless
steel crossed tube configuration with a flexible bellow
construction and viewports on each end of the 30~cm horizontal
arm. Using a thermocoax element the centre part of the vapour cell
is heated to 350~$^{\circ}$C yielding a maximum unsaturated
absorption of about 20$\%$. At this temperature the Doppler width
is 3.5~GHz. Lithium vapour is limited by its mean free path to the
centre part of the tube, protecting the viewports from becoming
coated with lithium. To further reduce this problem we heat the
viewports to 120~$^{\circ}$C.

\begin{figure}[htb]
\begin{center}
\epsfig{file=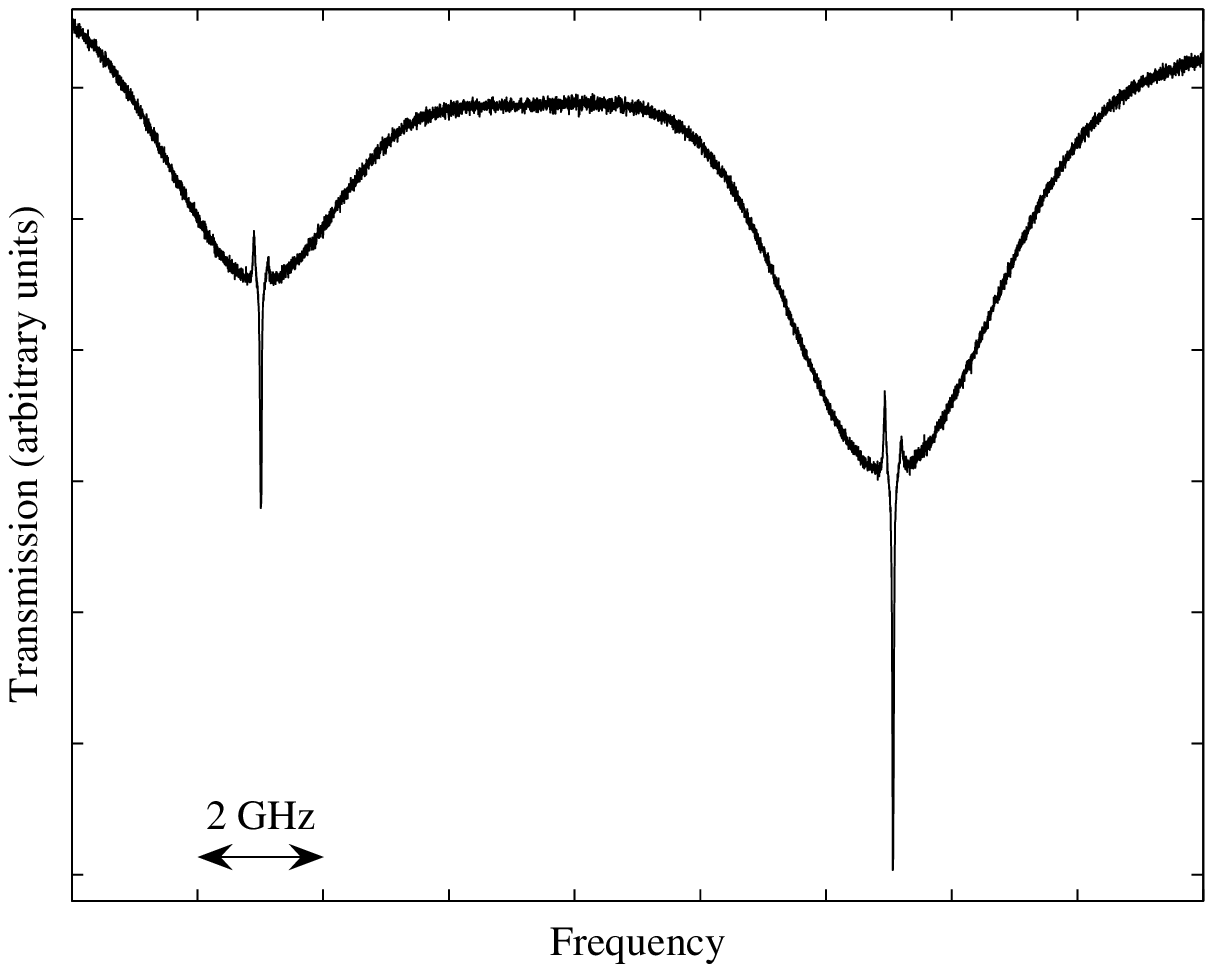,width=0.8\linewidth}

\vspace{-3.95cm}
\includegraphics[width=0.3\linewidth]{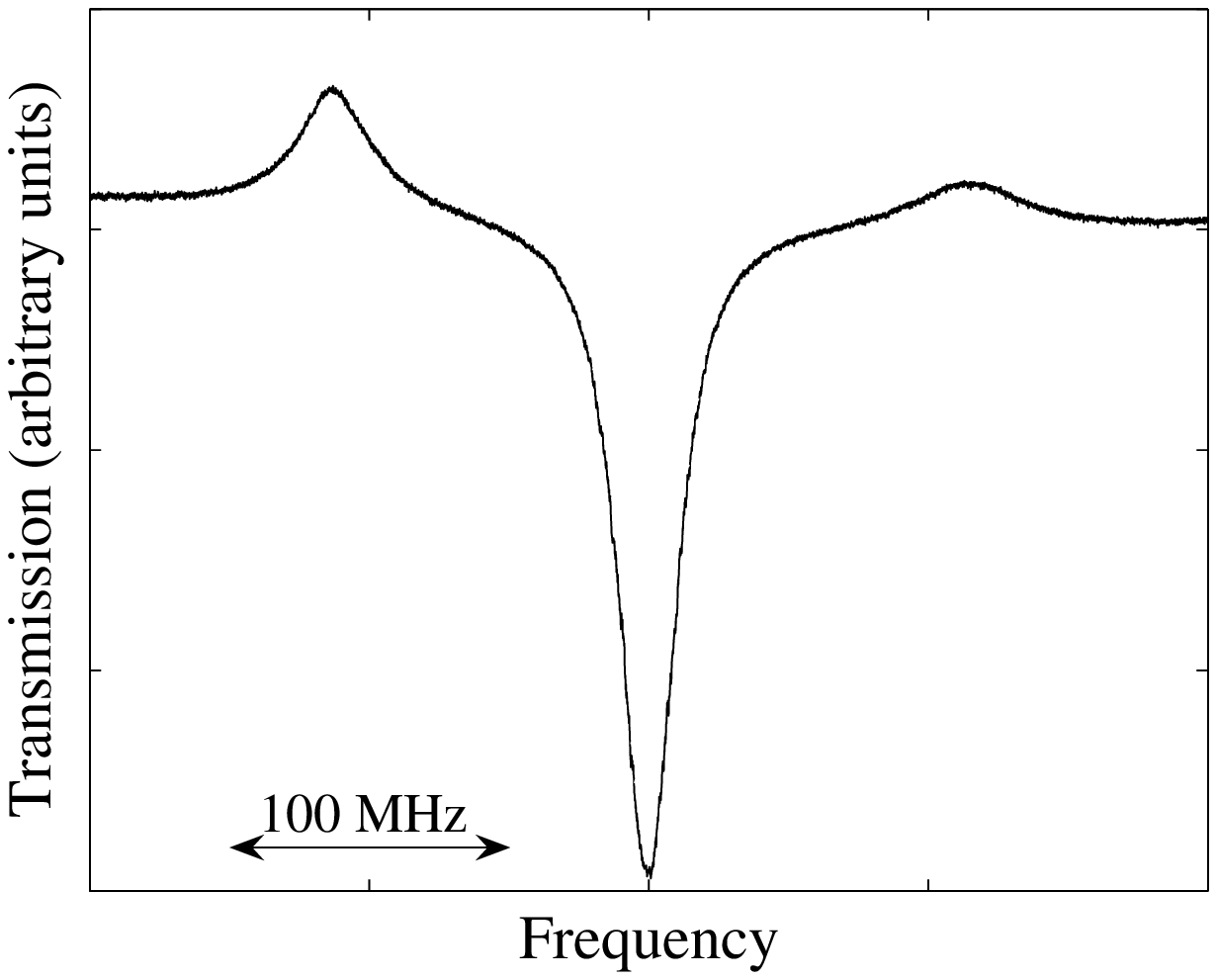}

\end{center}
\caption{Doppler-free spectrum of the D1 (a) and D2 (b) lines in
$^{6}$Li. The inset shows a close-up of the scan across the D2
line. The outer two peaks in the inset correspond to F~=~3/2
$\rightarrow$ F$^{'}$~=~5/2, 3/2, 1/2 and F~=~1/2 $\rightarrow$
F$^{'}$~=~3/2, 1/2 transitions.} \label{fig:sat}

\begin{picture}(0,0)
\put(110,220){\scriptsize(a)} \put(237,220){\scriptsize(b)}

\end{picture}
\end{figure}

Saturation spectroscopy of lithium vapour provides an excellent
frequency reference for our experiments in an atomic beam.
Figure~\ref{fig:sat} shows typical Doppler-free spectra of
$^{6}$Li with a frequency scan over both the D1
(figure~\ref{fig:sat}~(a)) and D2 (figure~\ref{fig:sat}~(b))
lines. The inset shows a close up of the D2 line whose outer peaks
correspond to the F~=~3/2 $\rightarrow$ F$^{'}$~=~5/2, 3/2, 1/2
and F~=~1/2 $\rightarrow$ F$^{'}$~=~3/2, 1/2 transitions,
respectively. The reduction in absorption of the probe at these
outer peaks is due to the saturation of the transition by the pump
beam. The enhancement in absorption of the crossover dip is due to
compensation of optical hyperfine pumping, which occurs when the
pump and probe laser are resonant with both ground state sublevels
at the same time. When applying higher intensities the
spectroscopy of the D1 line resembles the spectroscopy of the D2
line (inset in figure~\ref{fig:sat}). The width of the crossover
peak in the D2 line is $\approx$ 25~MHz. The contrast ratio of the
crossover peak with respect to the Doppler broadened resonance is
$\approx$ 100$\%$.

\begin{figure}[htb]
\begin{center} \psfrag{ECDL}{\tiny{ECDL}} \psfrag{Slave 1}{\tiny{Slave 1}}
\psfrag{Slave 2}{\tiny{Slave 2}} \psfrag{Slave 3}{\tiny{Slave 3}}
\psfrag{(pump)}{\tiny{(pump)}} \psfrag{(probe)}{\tiny{(probe)}}
\psfrag{sat}{\hspace{2mm}\tiny{saturation}}
\psfrag{vap}{\hspace{2mm}\tiny{vapour cell/}}
\psfrag{spec}{\hspace{2mm}\tiny{spectroscopy}}
\psfrag{at}{\hspace{2mm}\tiny{atomic beam}}
\psfrag{lam4}{{\tiny$\lambda$/4}}
\psfrag{lam2}{{\tiny$\lambda$/2}}
\psfrag{x}{\raisebox{1.5mm}{\hspace{4mm}\tiny-(90$\rightarrow$110)}}
\psfrag{xx}{\raisebox{1mm}{\hspace{7mm}\tiny MHz}}
\psfrag{y}{\raisebox{0mm}{\hspace{5mm}\tiny-114~MHz}}
\psfrag{z}{\raisebox{0mm}{\hspace{5mm}\tiny+87~MHz}}
\epsfig{file=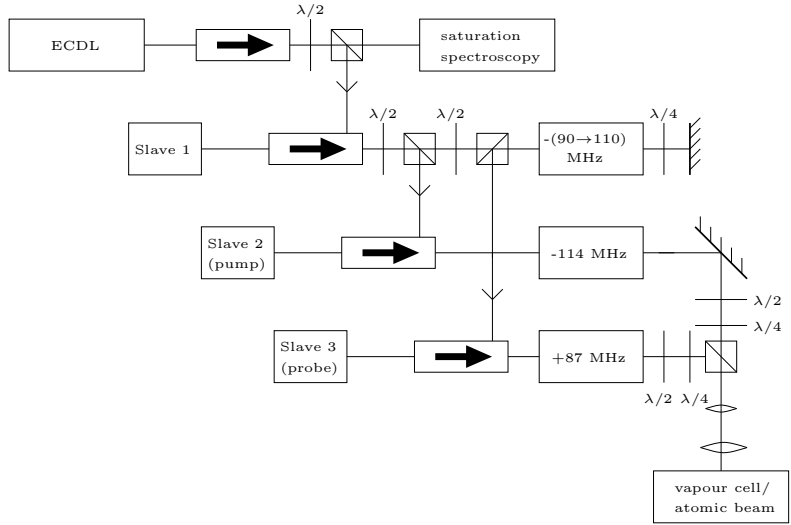,width=0.8\linewidth}
\caption{Experimental set-up for the EIT experiments in both the
vapour cell and the atomic beam. The ECDL is tuned to either the
D1 or D2 line. Three slave lasers were injection locked to the
master laser. Where necessary, the frequencies were shifted using
AOMs. The frequency difference sweep of the pump and probe laser
was generated by the AOM in double pass. The typical frequencies
used in the AOMs are shown in the above figure. Note that the
saturation spectroscopy is only used in conjunction with the
atomic beam experiment.} \label{fig:speccell}
\end{center}
\end{figure}

Figure~\ref{fig:speccell} shows the experimental set-up used for
EIT experiments in both the vapour cell and atomic beam. The EIT
resonances are usually observed with a bichromatic laser beam
consisting of two components with tunable frequency offset
changing in the vicinity of the ground state hyperfine splitting.
The ECDL (master laser) is used to injection lock a diode laser
(slave 1) which injection locks two additional diode lasers (slave
2 and slave 3). The master laser is tuned to either the D1 or D2
line. For the vapour cell experiments the master laser is not
actively stabilized since its long-term drift is small compared to
the Doppler width. However for the atomic beam experiments the
master laser is actively locked to the crossover of the saturated
absorption lines. Radiation of slave 3 is used as a pump beam for
coupling states between the ground state $F~=~1/2$ and the excited
levels in the D1 or D2 line. Laser light from slave 2 which is
resonant with the F~=~3/2~$\rightarrow$~F$'$ transitions (either
D1 or D2 line) is used as a probe beam. The probe laser frequency
is scanned over $\approx$ 25~MHz while the frequency of the pump
laser is fixed. They are overlapped on a non-polarizing beam
splitting cube and expanded to a beam diameter of 5~mm for the
vapour cell and 18~mm for the atomic beam experiments.

\section{Vapour Cell EIT}

By tailoring the optical frequencies present we performed
experiments on EIT in both a vapour cell and an atomic beam. To
obtain hyperfine ground state coherence we use mutually coherent
laser fields whose frequencies are separated by the hyperfine
splitting of the ground state. The pump and probe are sent through
the vapour cell with intensities of 8~mW/cm$^{2}$ and
2~mW/cm$^{2}$ respectively. To improve the signal to noise ratio
the pump is amplitude modulated by means of an mechanical chopper
(1~kHz) and the probe is detected using a lock-in amplifier. For
the results shown here, both beams are linearly polarized and
orthogonal to each other.

\vspace{5mm}
\begin{figure}[htb]
\begin{center}
\begin{minipage}{0.32\linewidth}
\includegraphics[width=1\linewidth,height=0.8\linewidth]{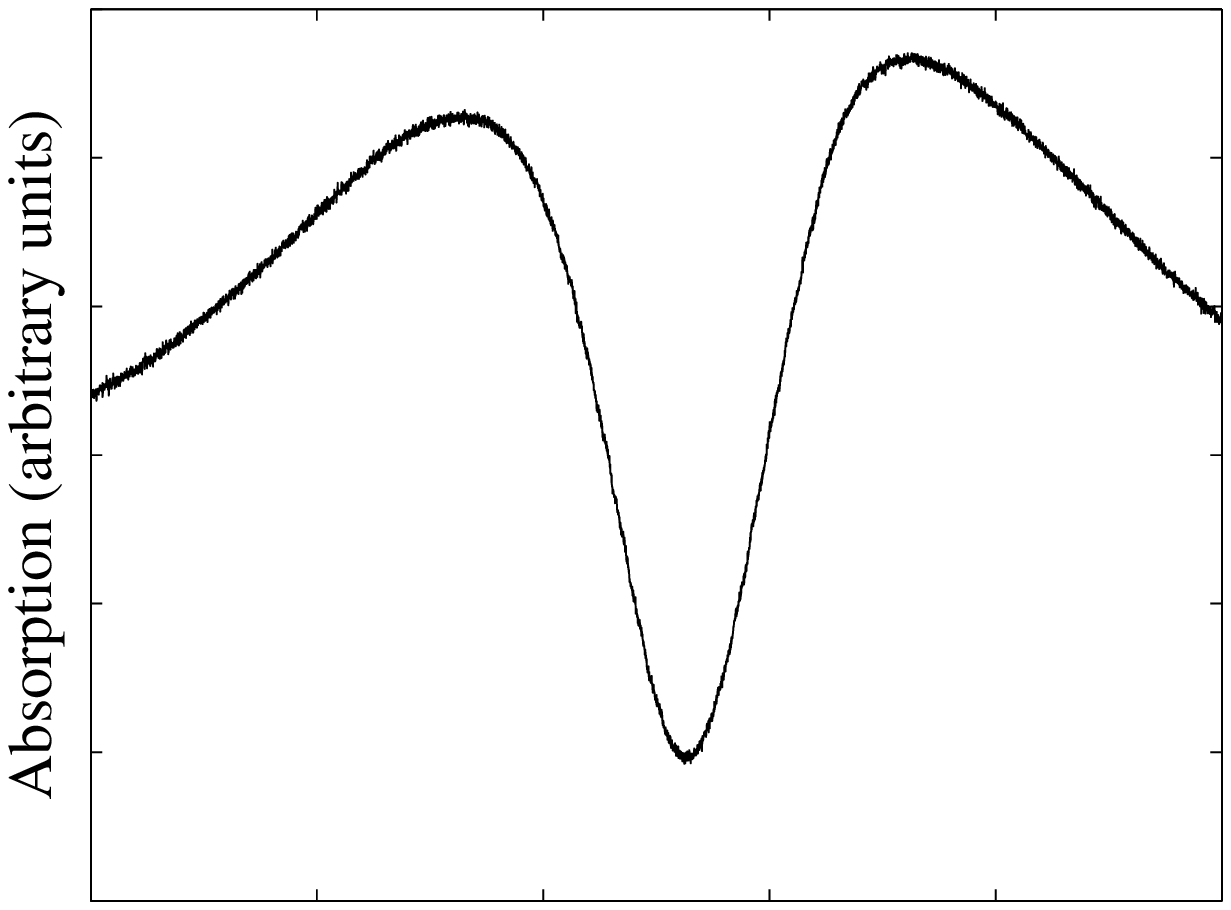}
\end{minipage}
\begin{minipage}{0.32\linewidth}
\includegraphics[width=1\linewidth,height=0.8\linewidth]{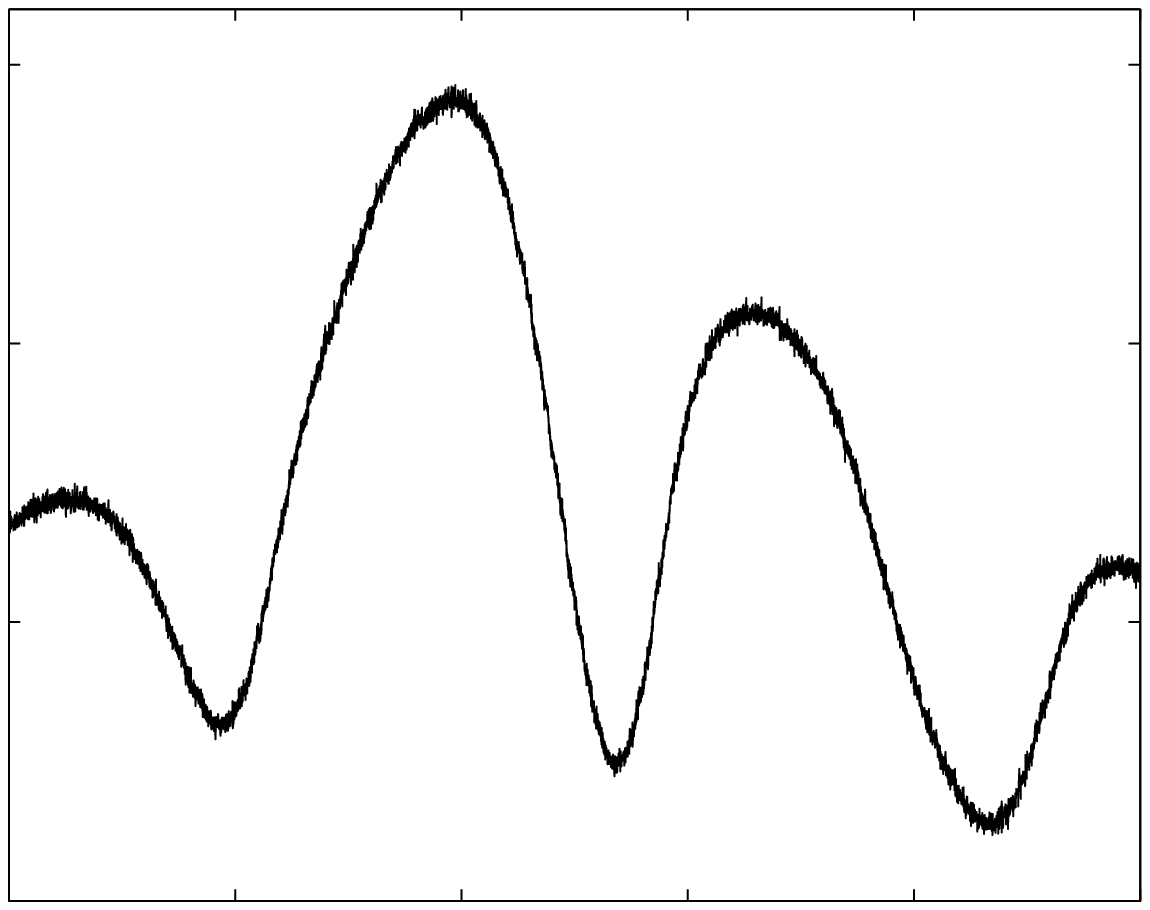}
\end{minipage}
\begin{minipage}{0.32\linewidth}
\emph{ }
\end{minipage}
\begin{minipage}{0.32\linewidth}
\includegraphics[width=1\linewidth,height=0.8\linewidth]{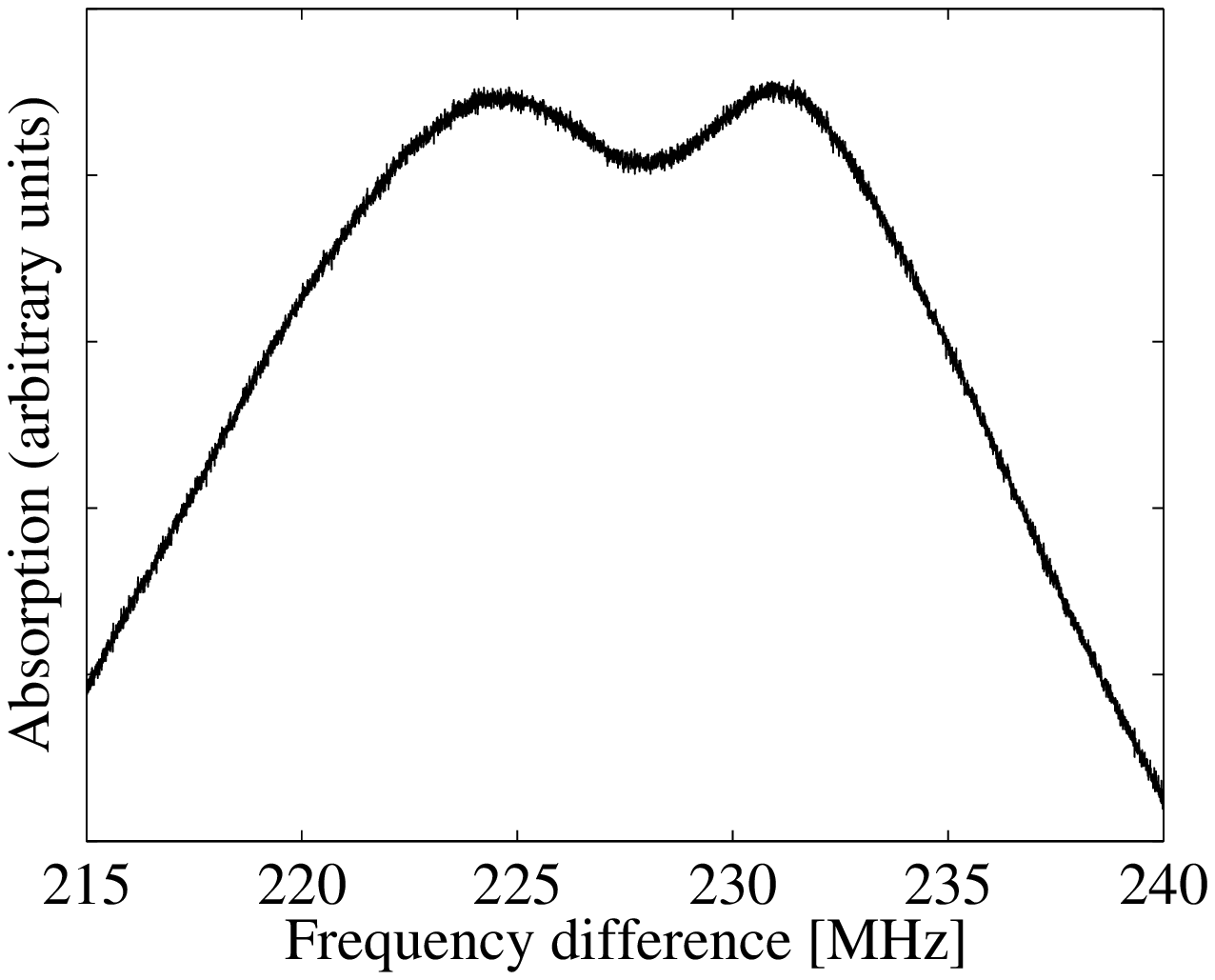}
\end{minipage}
\begin{minipage}{0.32\linewidth}
\includegraphics[width=1\linewidth,height=0.8\linewidth]{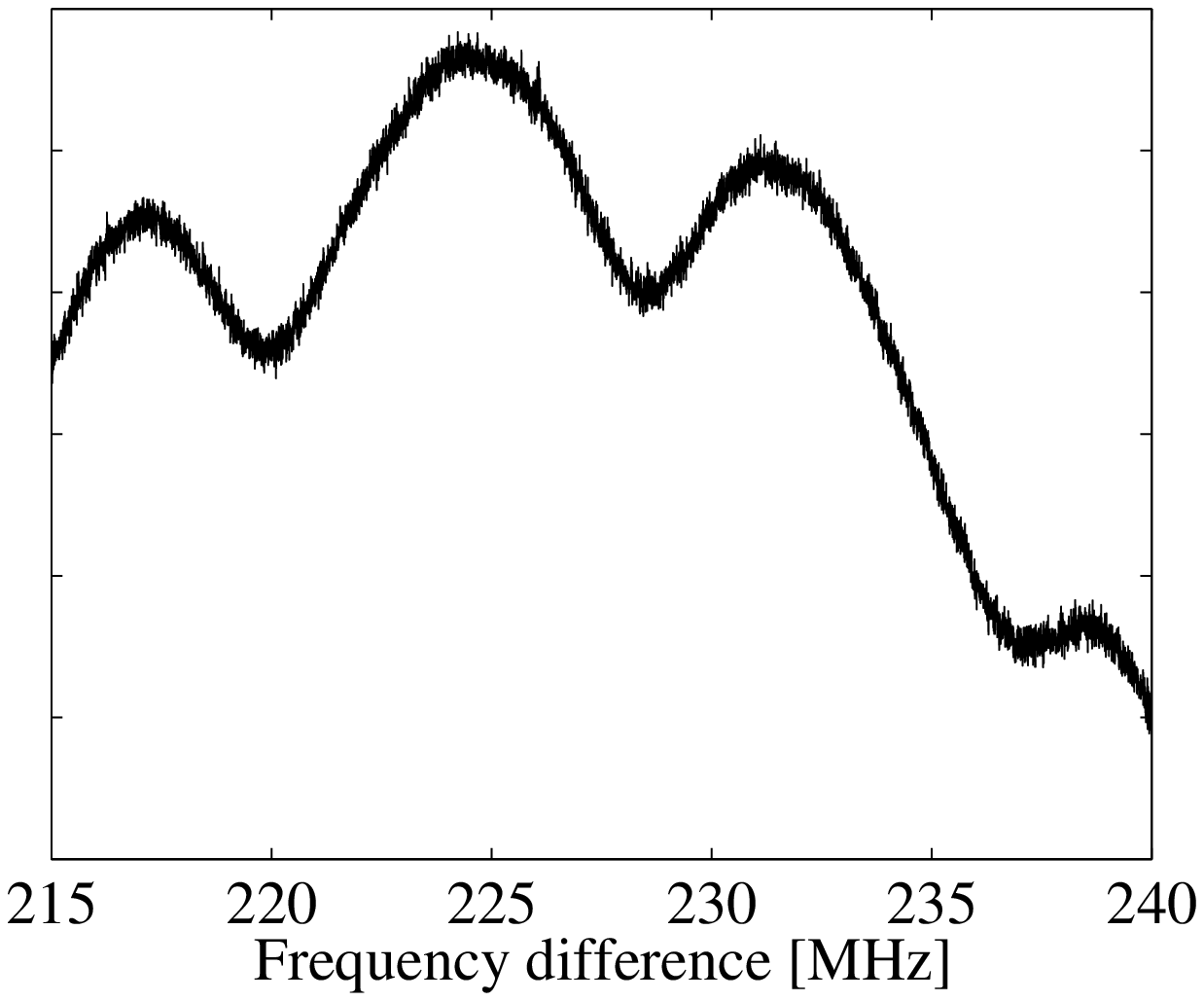}
\end{minipage}
\begin{minipage}{0.32\linewidth}\vspace{-30mm}
\includegraphics[width=0.6\linewidth]{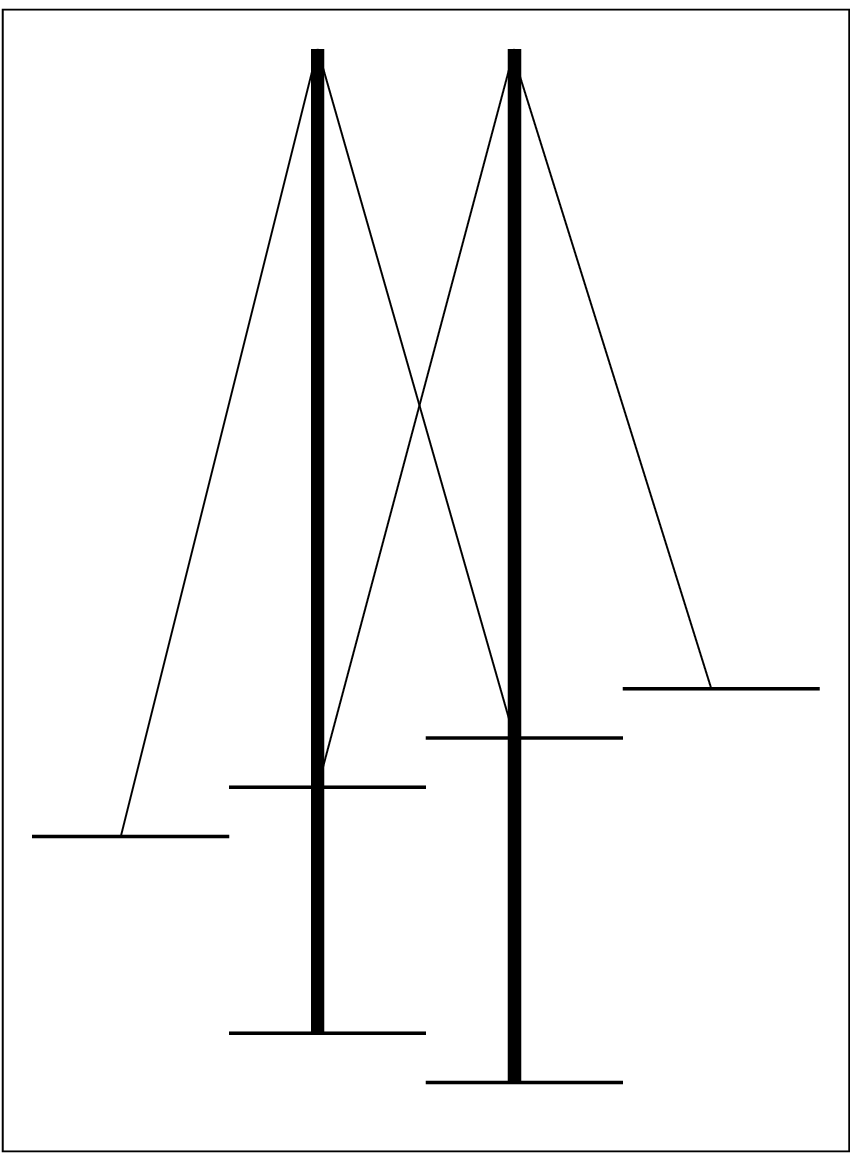}
\end{minipage}
\end{center}
\caption{Absorption of the D1 (a and b) and D2 (c and d) line as a
function of frequency difference between the pump and probe. (a)
and (c): B~=~0, (b) and (d) have a magnetic field of $\approx$~5~G
perpendicular to the laser light. (e) shows the Raman transitions
responsible for coupling the ground state Zeeman sublevels for
each of the resonances in (b and d). The thicker lines represent
the pump laser which is tuned to the F~=~1/2~$\rightarrow$~F$^{'}$
transitions.} \label{fig:EITsc}
\begin{picture}(0,0)
\put(22,257){\scriptsize(a)} \put(22,161){\scriptsize(c)}
\put(144,257){\scriptsize(b)} \put(144,161){\scriptsize(d)}
\put(252,211){\scriptsize(e)}

\put(-2,230){D1} \put(-2,130){D2} \put(57,272){B=0}

\put(179,272){$\vec{B}\bot\vec{k}$}
\end{picture}
\end{figure}

\subsection{Results}

Figure~\ref{fig:EITsc} shows absorption plots of the D1 and D2
lines as a function of frequency difference between the collinear
pump and probe beams. With no externally applied magnetic field a
single absorption dip at a frequency difference of 228~MHz is
observed (figure~\ref{fig:EITsc}~(a) and \ref{fig:EITsc}~(c)). The
width of these resonances is approximately 4~MHz which is less
than the natural width of 5.9~MHz. The amplitudes for the D2 line
are notably weaker than for the D1 line. This is due to
destructive excitations via cycling transitions which do not
contribute to the preparation of dark coherent states. This is in
agreement with the results of St\"ahler~{\emph et~al}
\cite{Stahler02} and Magnus~{\emph et~al} \cite{Magnus05} which
show greater contrast in the dark resonance in the D1 line than
the D2 line of $^{85}$Rb and $^7$Li, respectively.

It is difficult to compare the width of the EIT resonances
obtained for both lines because any ambient magnetic field can
introduce additional broadening. To investigate the possible
ambient field broadening an external homogeneous magnetic field
(approximately 5 G) was applied perpendicular to the laser beam
{$\vec{B}\bot\vec{k}$}. The magnetic field removes the degeneracy
of the Zeeman levels, splitting the sub-natural EIT resonance
(figure~\ref{fig:EITsc}~(b and d)). The applied magnetic field
also introduces a convenient quantization axis. The pump light
with linear polarization parallel to the magnetic field can
produce $\pi$ transitions, while the probe with orthogonal linear
polarization excites the $\sigma^{\pm}$ transitions
(figure~\ref{fig:EITsc}~(e)). The m~=~0~$\rightarrow$~m${'}$~=~0
type Raman transition, which is magnetic field insensitive in the
linear Zeeman approximation, does not exist for $^6$Li, however,
the Raman transitions
F~=~1/2~(m~=~1/2)~$\rightarrow$~F$^{'}$~=~3/2~(m$^{'}$~=~-1/2) and
F~=~1/2~(m~=~-1/2)~$\rightarrow$~F$^{'}$~=~1/2~(m$^{'}$~=~1/2) are
also magnetic field insensitive, because the Zeeman shift of the
upper and lower magnetic sub-levels are almost equal. Thus, the
EIT dip in absorption, which remains unshifted at 228~MHz
frequency difference with increasing magnetic field is due to the
above mentioned Raman transitions. The width of this EIT resonance
for the D1 line is approximately 3~MHz. Both outer dips are
symmetrically shifted by 2$\Delta$ from the unshifted centre dip,
where $\Delta = \mu_{B}g_{F}B/h$, $\mu_{B}$ is the Bohr magneton,
$h$ is Planck's constant and $g_{F}$ is the Land\'e factor. The
width of the outer dips are $25\%$ larger than the width of the
unshifted dip due to spatial inhomogeneities in the magnetic
field.

We reduced the intensities of the pump and probe lasers but
observed no change in the width of the EIT resonances which
implies that the width is not limited by power broadening.
Although Li-Li collisions are negligible, collisions with residual
background gases may cause spin depolarizing collisions of the
ground states which possibly contribute to the EIT width. The
resolution is further limited by finite interaction (transit)
time. The spectroscopy of atoms in a metal vapour cell has the
limitations of randomly directed velocities (i.e. Doppler
broadening), field inhomogeneities and high collision rates.

\section{Atomic Beam EIT}

To improve the resolution of EIT resonances experiments were
performed using the standard technique of a collimated atomic
beam. Isotopically enriched $^{6}$Li atoms are evaporated from an
oven at a temperature of 450$^\circ$~C to produce a thermal atomic
beam. At this temperature the $^{6}$Li vapour pressure in the oven
is expected to be 4$\times$10$^{-4}$~Torr. The diameter of the
beam at the centre of the probe region is 7~mm. Ultra high vacuum
of 2$\times$10$^{-10}$~Torr is achieved in an anti-reflection
coated glass cell (for 670~nm) which has outer dimensions of
3x3x12~cm$^3$ with a wall thickness of 3~mm. The collimated atomic
beam is excited at right angles by resonant laser light in the
interaction region. The pump and probe intensities were
1.4~mW/cm$^2$ for both beams. Fluorescence from atoms in the probe
region was detected using a photomultiplier tube. The cell was
wrapped in $\mu$-metal to reduce stray magnetic fields.

\subsection{Results}

\begin{figure}[htb]
\begin{center}
\begin{minipage}{0.49\linewidth}
\center
\includegraphics[width=1\linewidth,height=0.8\linewidth]{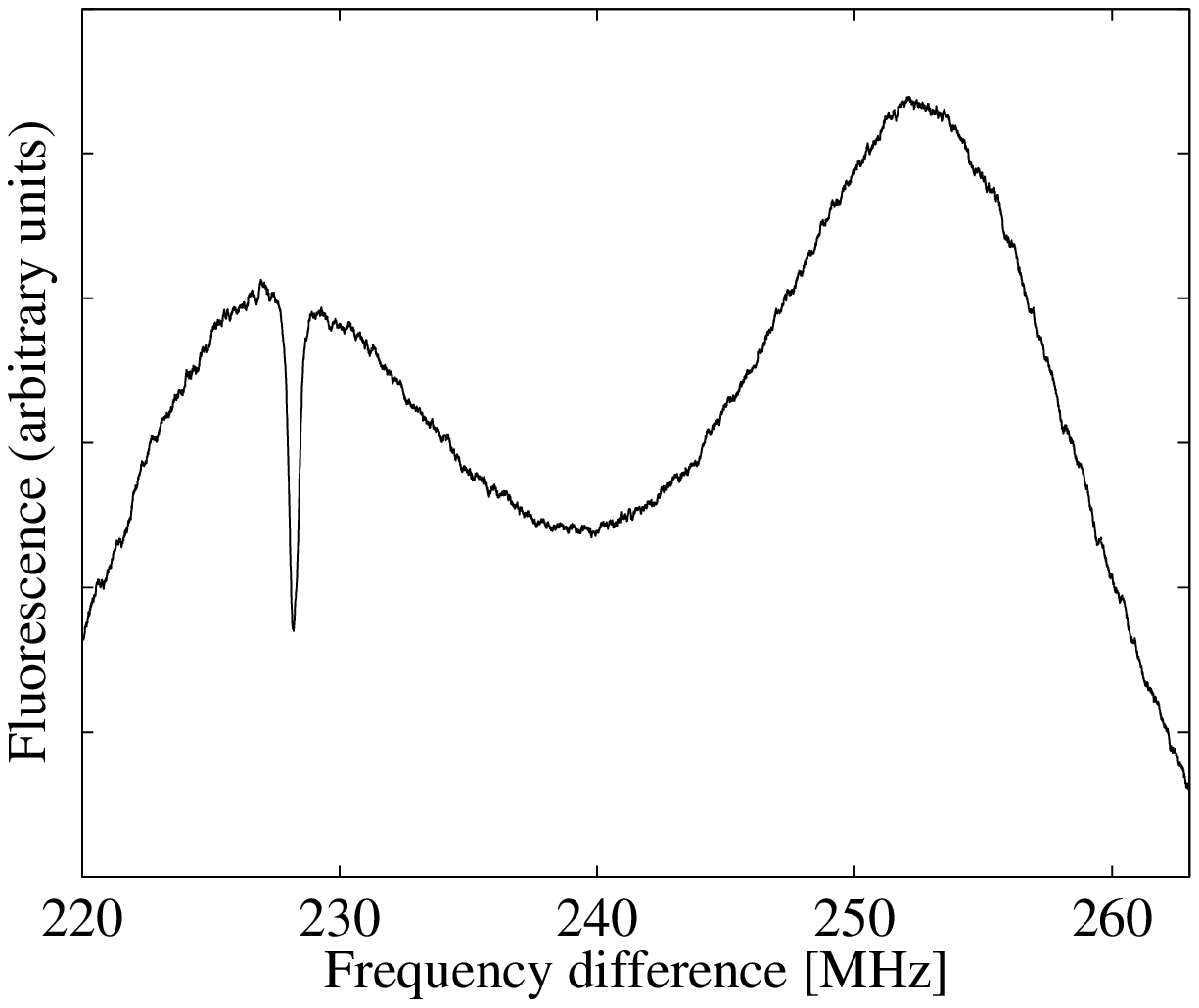}
\end{minipage}
\begin{minipage}{0.49\linewidth}
\includegraphics[width=1\linewidth,height=0.8\linewidth]{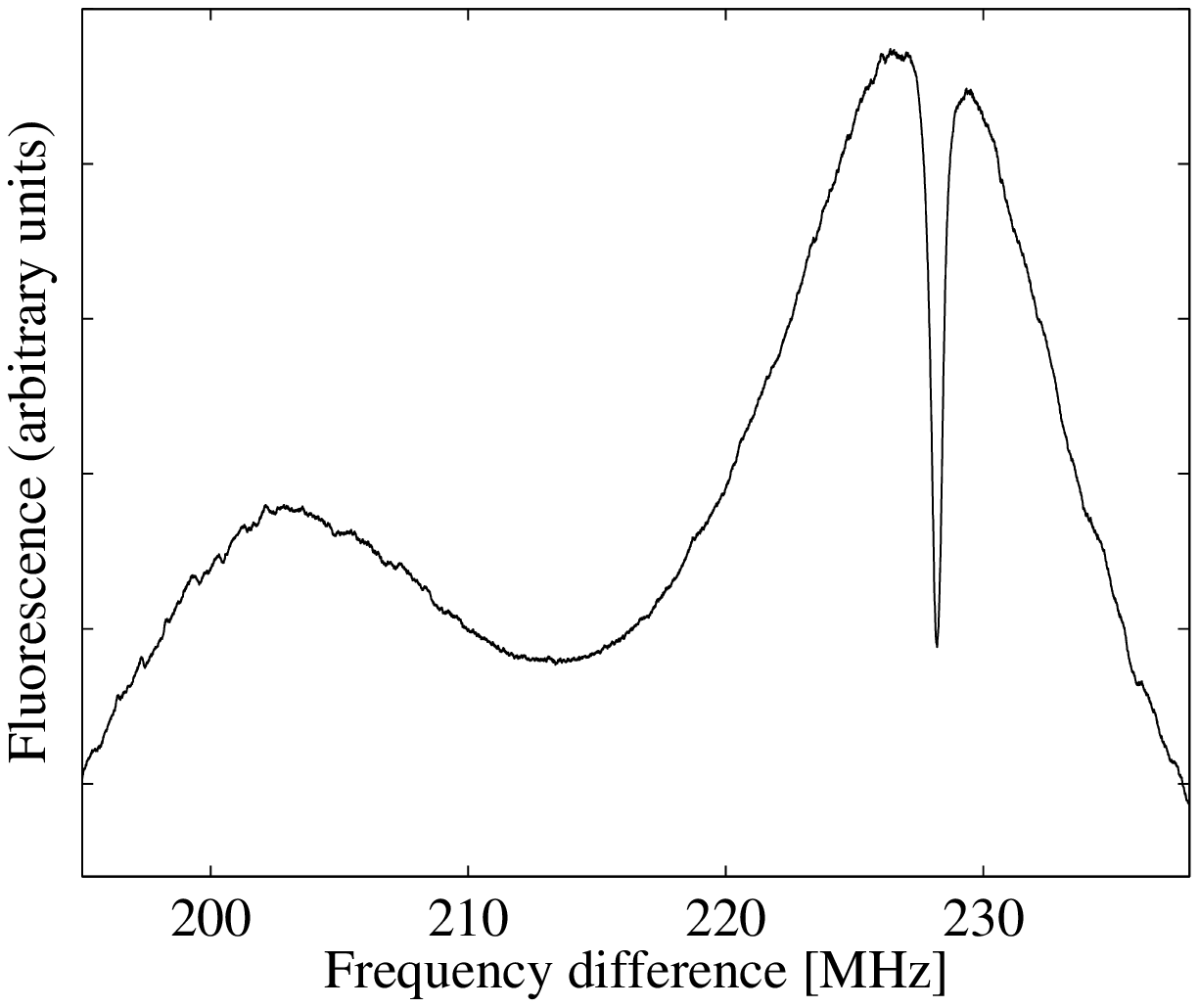}
\end{minipage}
\end{center}
\caption{D1 line fluorescence from $^6$Li atoms. Probe laser is
scanned over both the F~=~3/2~$\rightarrow$~F$^{'}$~=~1/2 and
F~=~3/2~$\rightarrow$~F$^{'}$~=~3/2 transitions. (a) Fixed pump
laser tuned to (a) F~=~1/2 $\rightarrow$ F$^{'}$~=~1/2 transition
and (b) F~=~1/2 $\rightarrow$ F$^{'}$~=~3/2 transition.}
\label{fig:EIT1232}
\begin{picture}(0,0)
\put(30,195){\scriptsize(a)} 
\put(215,195){\scriptsize(b)}
\end{picture}
\end{figure}

Resonant fluorescence in the collimated atomic beam showed a much
narrower transverse Doppler width of 20~MHz compared to $\approx$
3.5~GHz obtained in the vapour cell. Therefore, the hyperfine
splitting of the $2^2$P$_{1/2}$ state of 26.1~MHz can be resolved.
Figure~\ref{fig:EIT1232} shows a plot of fluorescence of the D1
line as a function of frequency difference between the pump and
probe laser. In this figure the probe laser is scanned over both
the F~=~3/2~$\rightarrow$~F$^{'}$~=~1/2 and
F~=~3/2~$\rightarrow$~F$^{'}$~=~3/2 transitions whereas the pump
laser has a fixed frequency tuned to the
F~=~1/2~$\rightarrow$~F$^{'}$~=~1/2 (a) or
F~=~1/2~$\rightarrow$~F$^{'}$~=~3/2 (b) transition. The EIT dip is
more pronounced when the fixed laser is tuned to the F$^{'}$~=~3/2
state which is due to the larger transition probability. For this
reason we performed all of the subsequent experiments with the
pump laser tuned to the F~=~1/2~$\rightarrow$~F$^{'}$~=~3/2
transition.

\begin{figure}[htb]
\begin{center}
\begin{minipage}{0.49\linewidth}
\center
\includegraphics[width=0.9\linewidth]{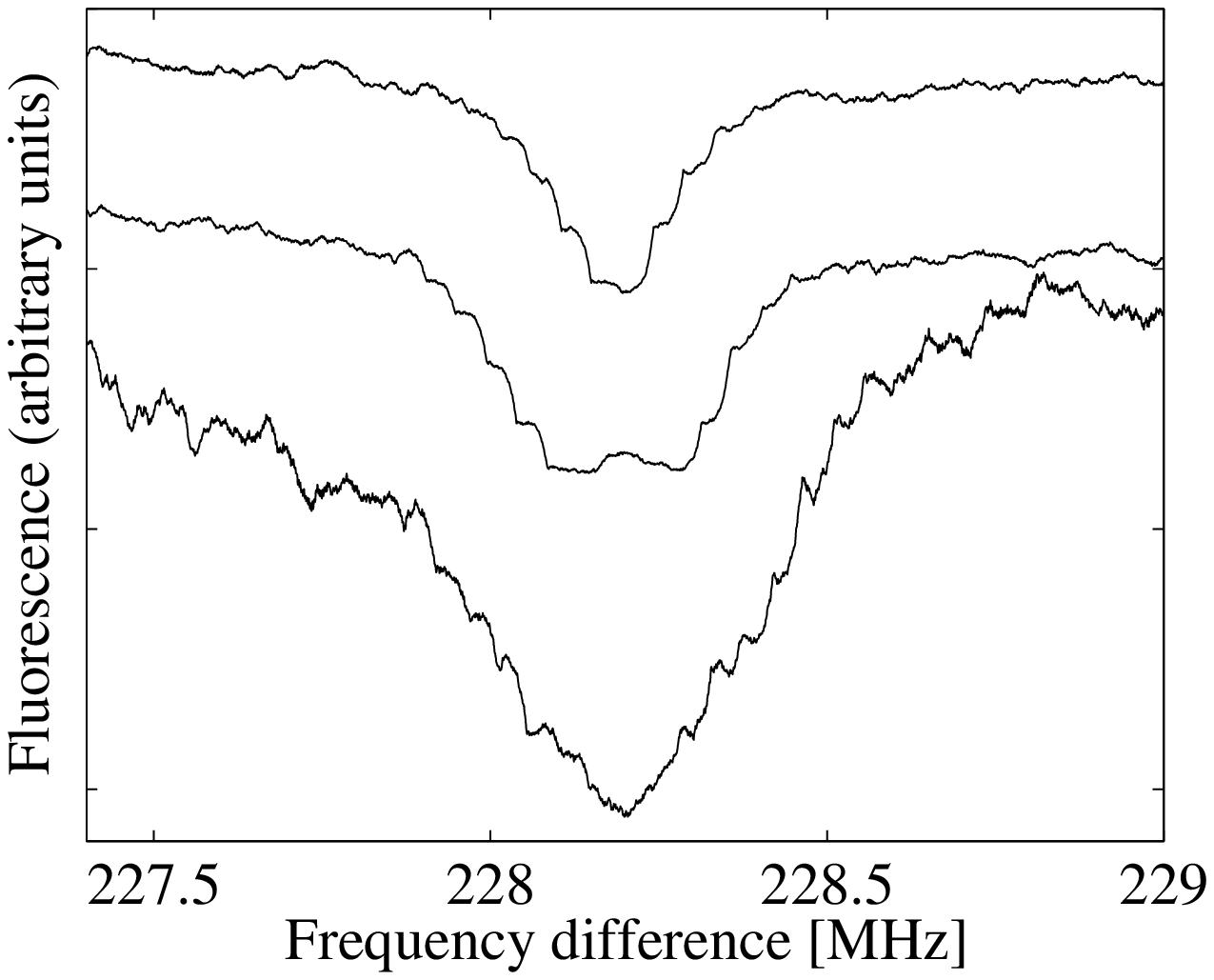}
\end{minipage}
\begin{minipage}{0.49\linewidth}
\psfrag{F=1/2}{\raisebox{0mm}{\tiny{F=1/2}}}
\psfrag{F=3/2}{\raisebox{0mm}{\hspace{-4mm}\tiny{F=3/2}}}
\psfrag{F'=1/2}{\raisebox{0mm}{\tiny{F$'$=1/2}}}
\psfrag{F'=3/2}{\raisebox{0mm}{\tiny{F$'$=3/2}}}
\psfrag{F'=5/2}{\raisebox{0mm}{\tiny{F$'$=5/2}}}
\includegraphics[width=0.48\linewidth]{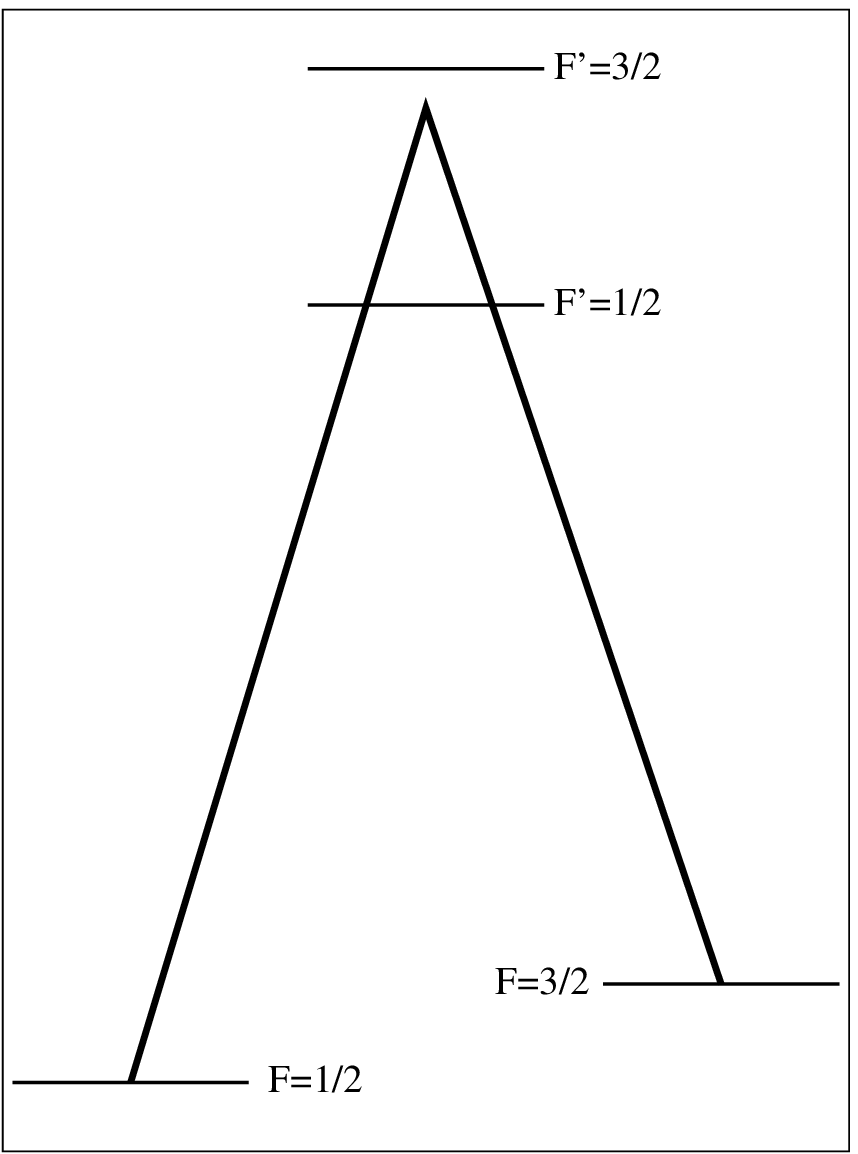}
\includegraphics[width=0.48\linewidth]{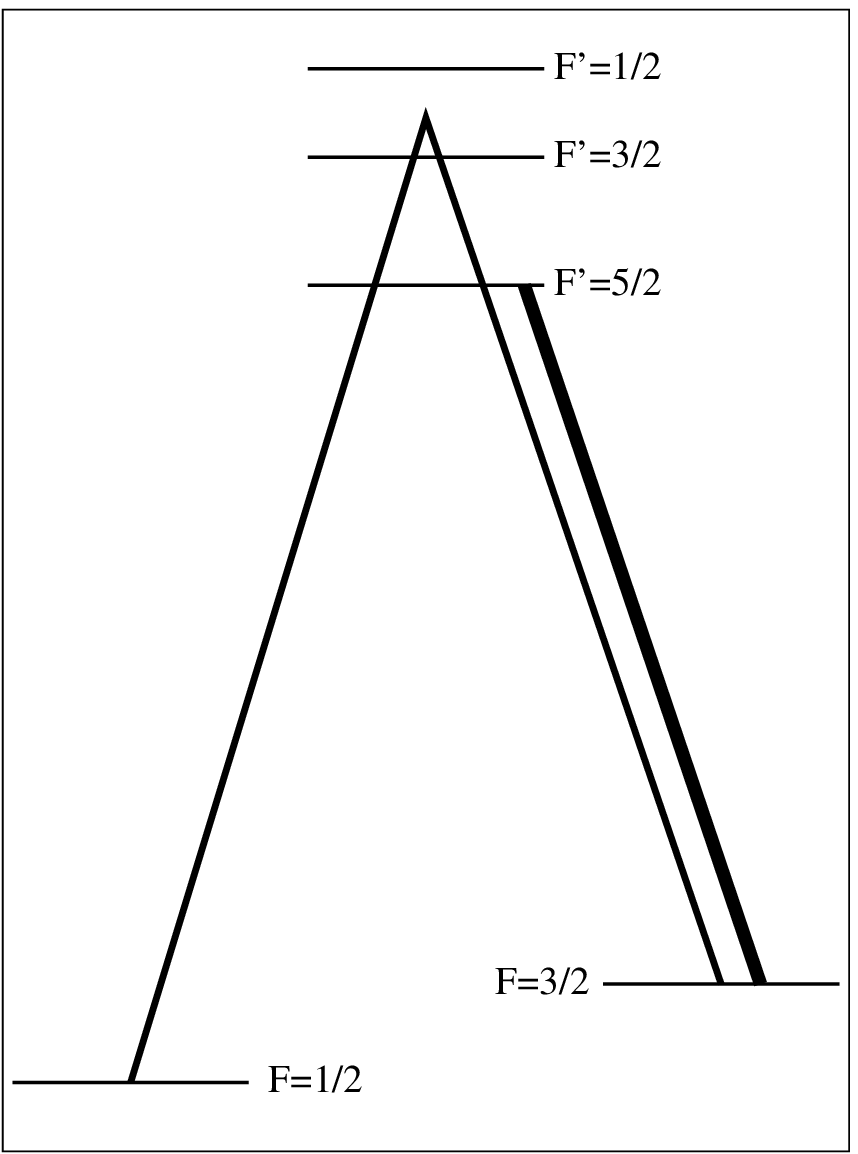}
\end{minipage}
\end{center}
\caption{Fluorescence of the D1 and D2 line as a function of
frequency difference between the pump and probe laser. (a) D1
resonance (width of 210~kHz) with applied magnetic field, (b) D1
resonance (370~kHz) with no magnetic field, (c) D2 resonance
(560~kHz) with no magnetic field. The Raman transition responsible
for the D1 and D2 resonance is shown in (d) and (e) respectively.}
\label{fig:width}
\begin{picture}(0,0)
\put(36,125){\scriptsize(c)} \put(36,150){\scriptsize(b)}
\put(36,170){\scriptsize(a)}
\put(190,185){\scriptsize(d)}\put(280,185){\scriptsize(e)}
\put(225,195){\scriptsize D1}\put(313,195){\scriptsize D2}
\end{picture}
\end{figure}

The EIT resonances obtained on the D1 and D2 lines are shown in
figure~\ref{fig:width}. The intensity of the resonant fluorescence
on the D2 line is higher, but the contrast of the sub-natural
resonances with respect to the residual Doppler background is
lower compared to the D1 line. In this regard the observations in
the atomic beam and the vapour cell are very similar. However,
better spectral resolution in the atomic beam allows us to
demonstrate that the width of EIT resonances in both cases is also
essentially different. The sub-natural resonances observed in the
atomic beam on both lines under very similar experimental
conditions such as light intensity, polarization and beam
overlapping are shown in figure~\ref{fig:width}. The curve in
figure~\ref{fig:width}~(c) represents EIT resonance observed on
the D2 line without applied magnetic field. The width of the
resonance is approximately 560~kHz. The sub-natural resonance on
the D1 line (370~kHz) (figure~\ref{fig:width}~(b)) reveals a
doublet structure due to residual magnetic field in the
interaction region. This splitting is hidden on the D2 line by the
large width of the EIT resonance. The double structure can be
removed by applying a small magnetic field along the atomic beam
resulting in the 210~kHz wide resonance
(figure~\ref{fig:width}~(a)) on the D1 line. We believe that the
EIT resonance on the D2 line is wider because of shorter lifetime
of the ground state coherence destructed via the cycling
transition F~=~3/2~$\rightarrow$~F$^{'}$~=~5/2. \\

\begin{figure}[h]
\begin{center}
\begin{minipage}{0.37\linewidth}
\includegraphics[width=1\linewidth,height=0.8\linewidth]{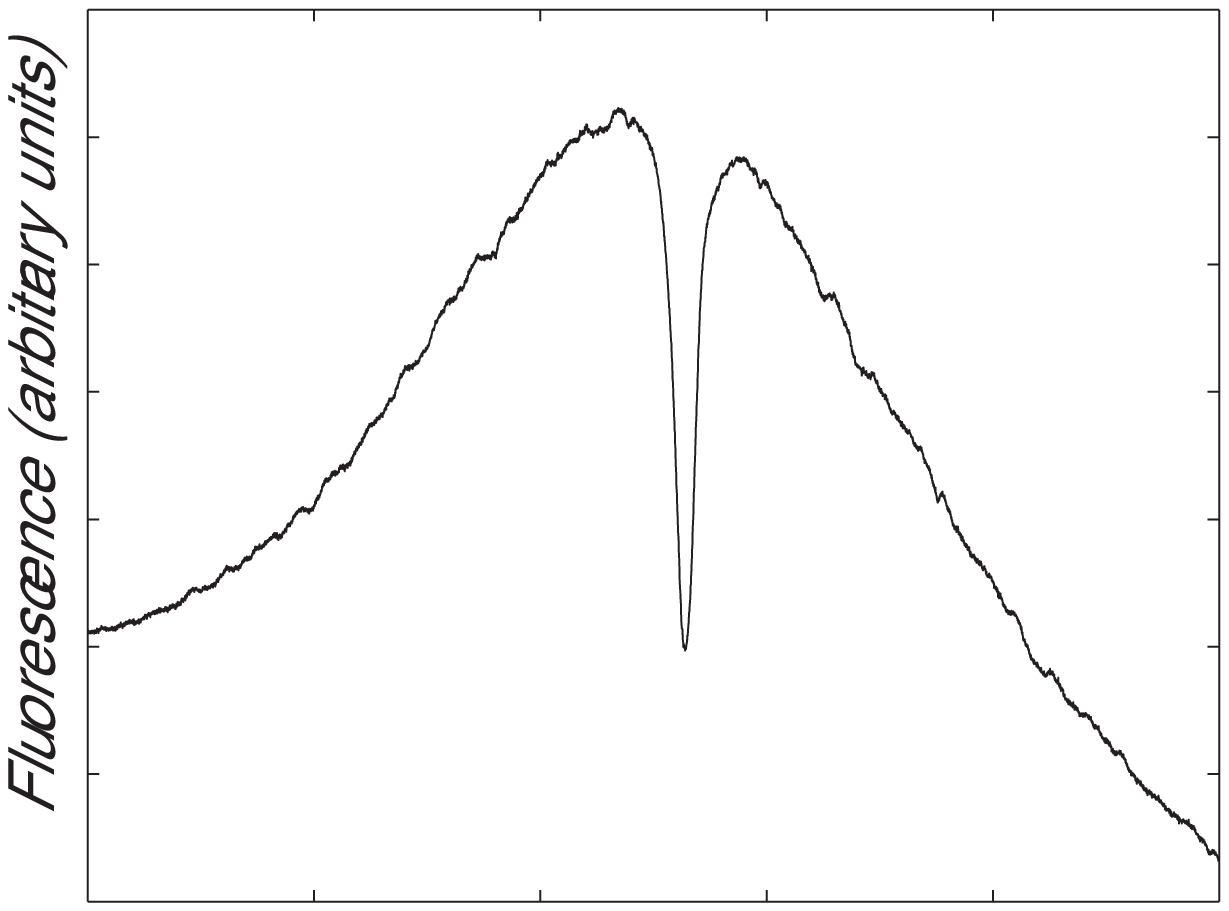}
\end{minipage}
\begin{minipage}{0.37\linewidth}
\includegraphics[width=1\linewidth,height=0.8\linewidth]{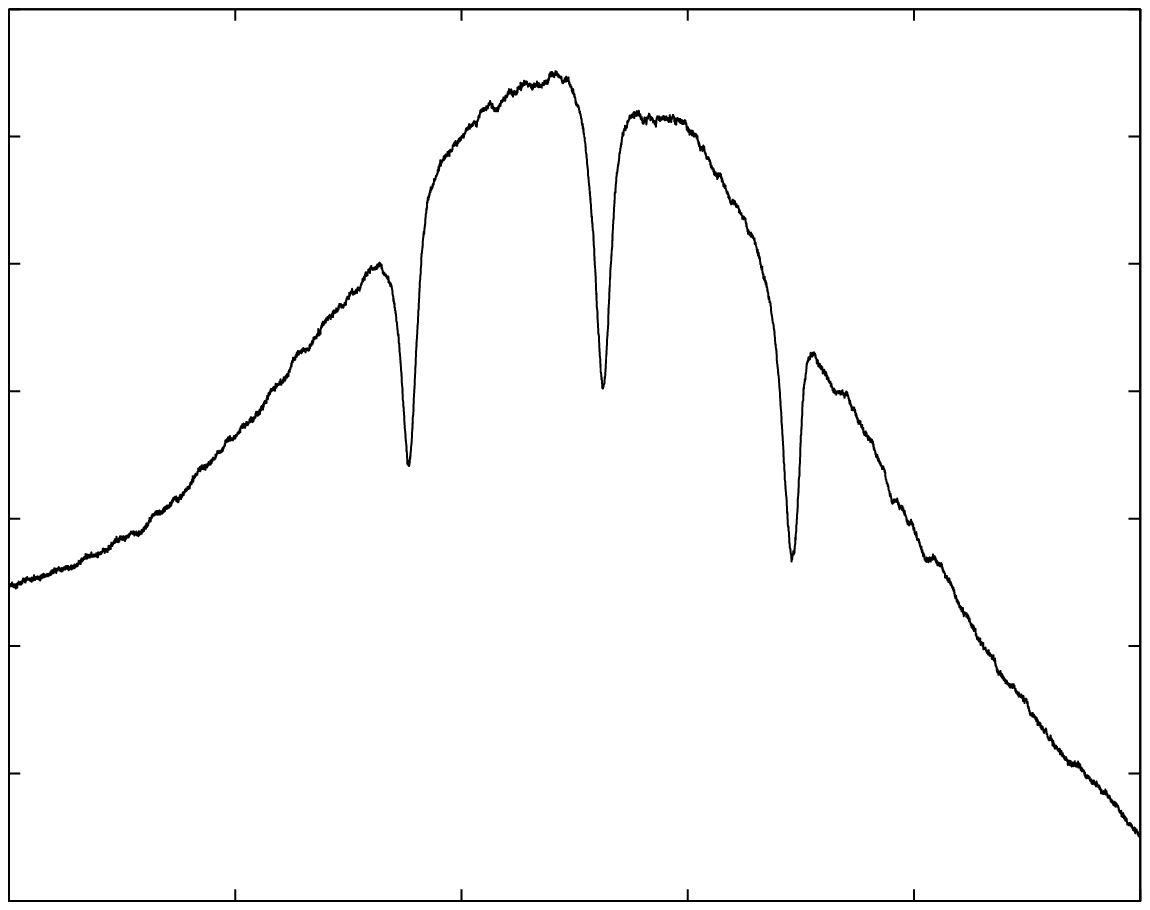}
\end{minipage}
\begin{minipage}{0.22\linewidth}\vspace{-2mm}
\includegraphics[height=1\linewidth]{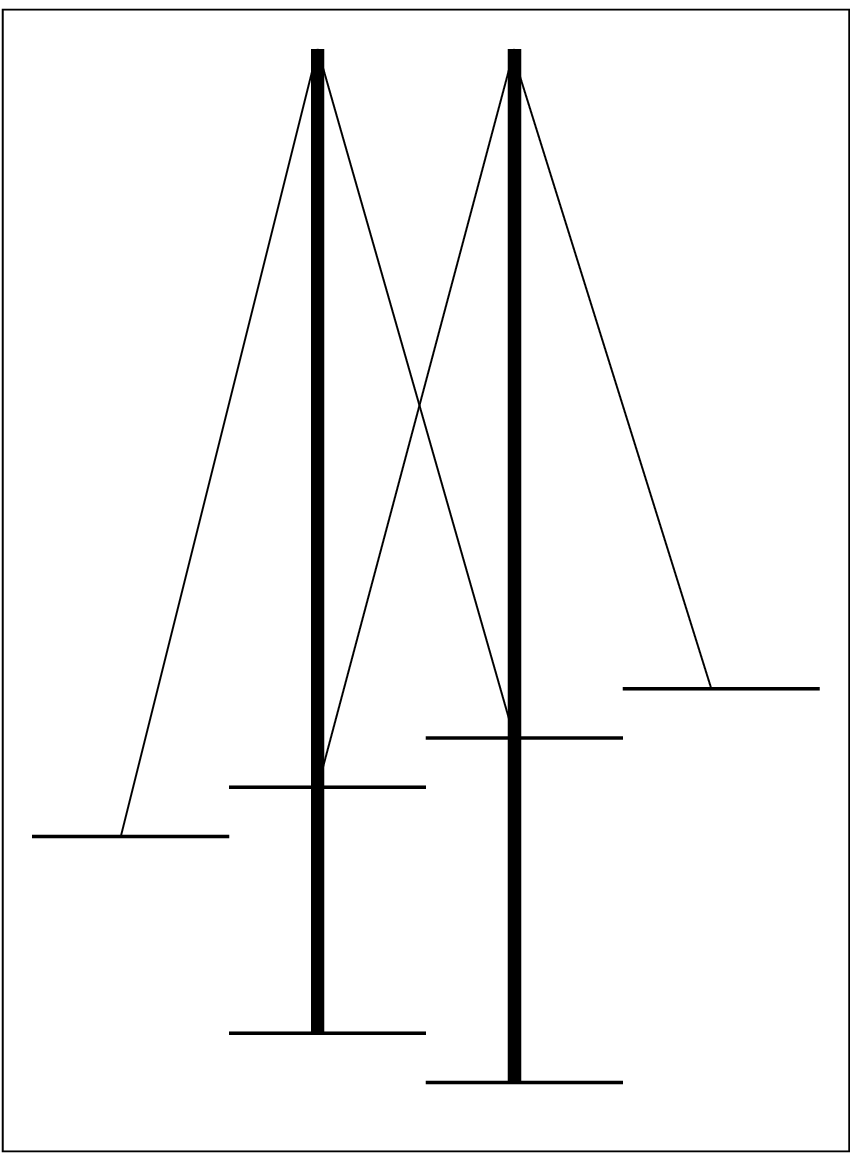}
\end{minipage}
\begin{minipage}{0.37\linewidth}
\includegraphics[width=1\linewidth,height=0.8\linewidth]{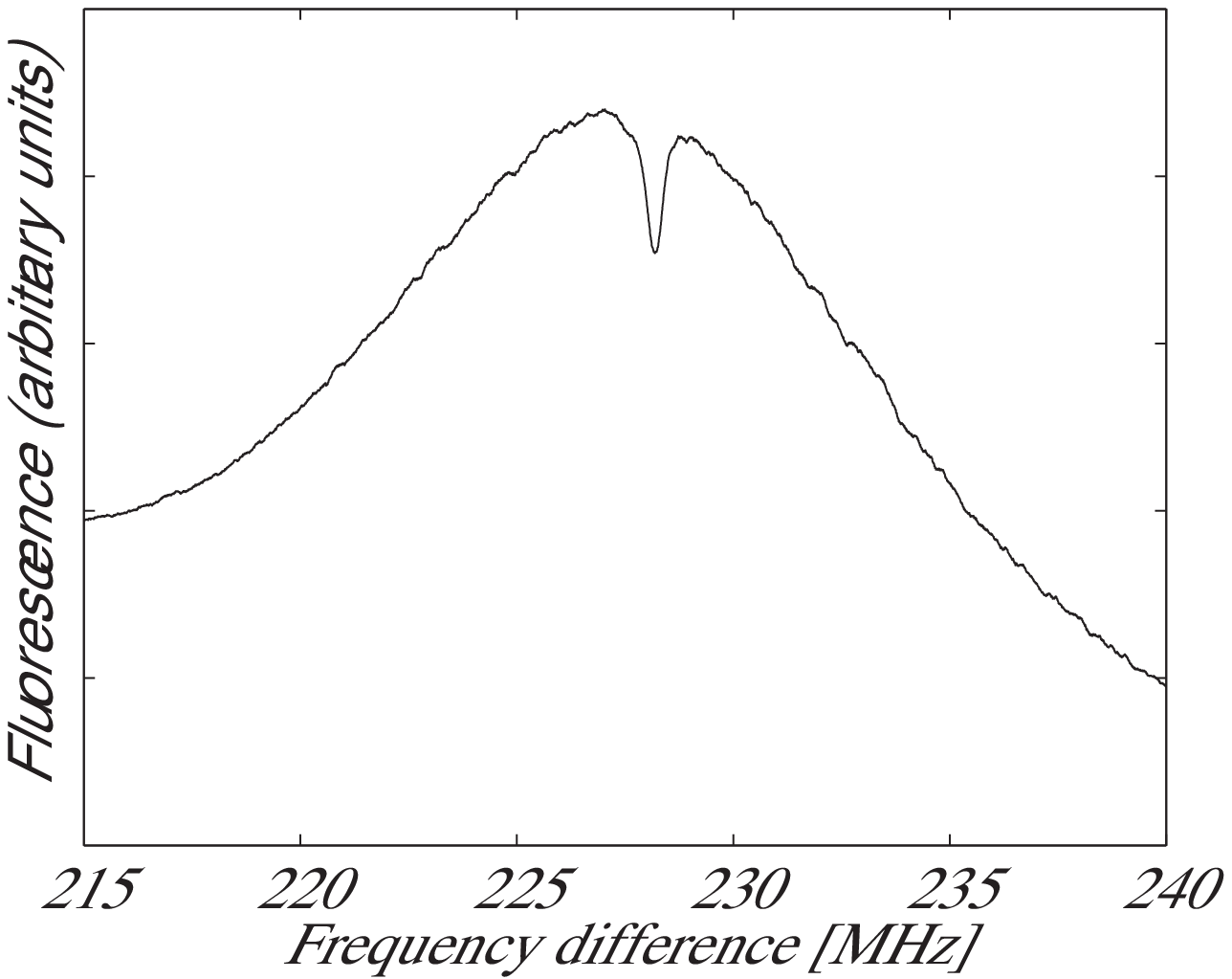}
\end{minipage}
\begin{minipage}{0.37\linewidth}
\includegraphics[width=1\linewidth,height=0.8\linewidth]{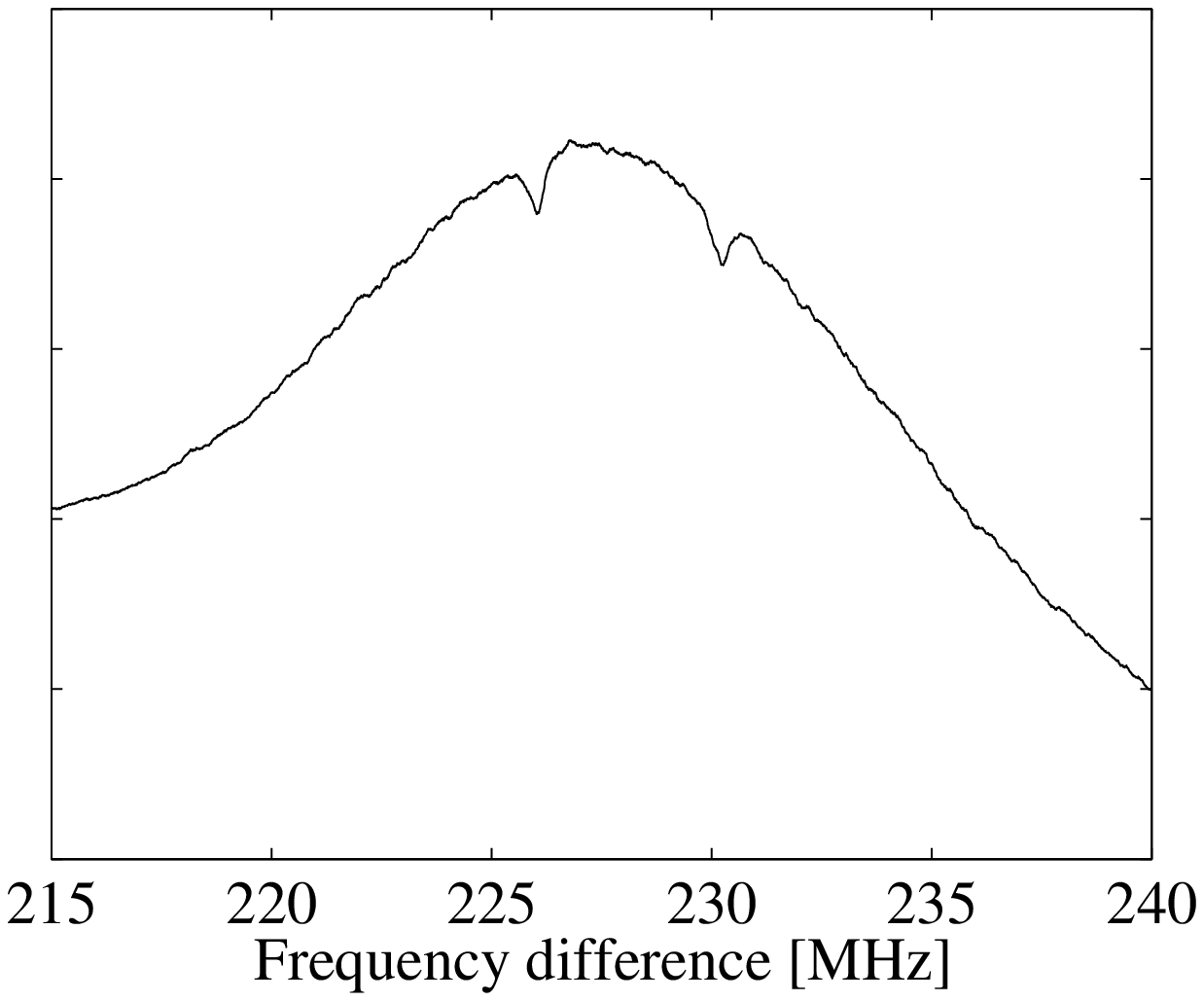}
\end{minipage}
\begin{minipage}{0.22\linewidth}\vspace{-2mm}
\includegraphics[height=1\linewidth]{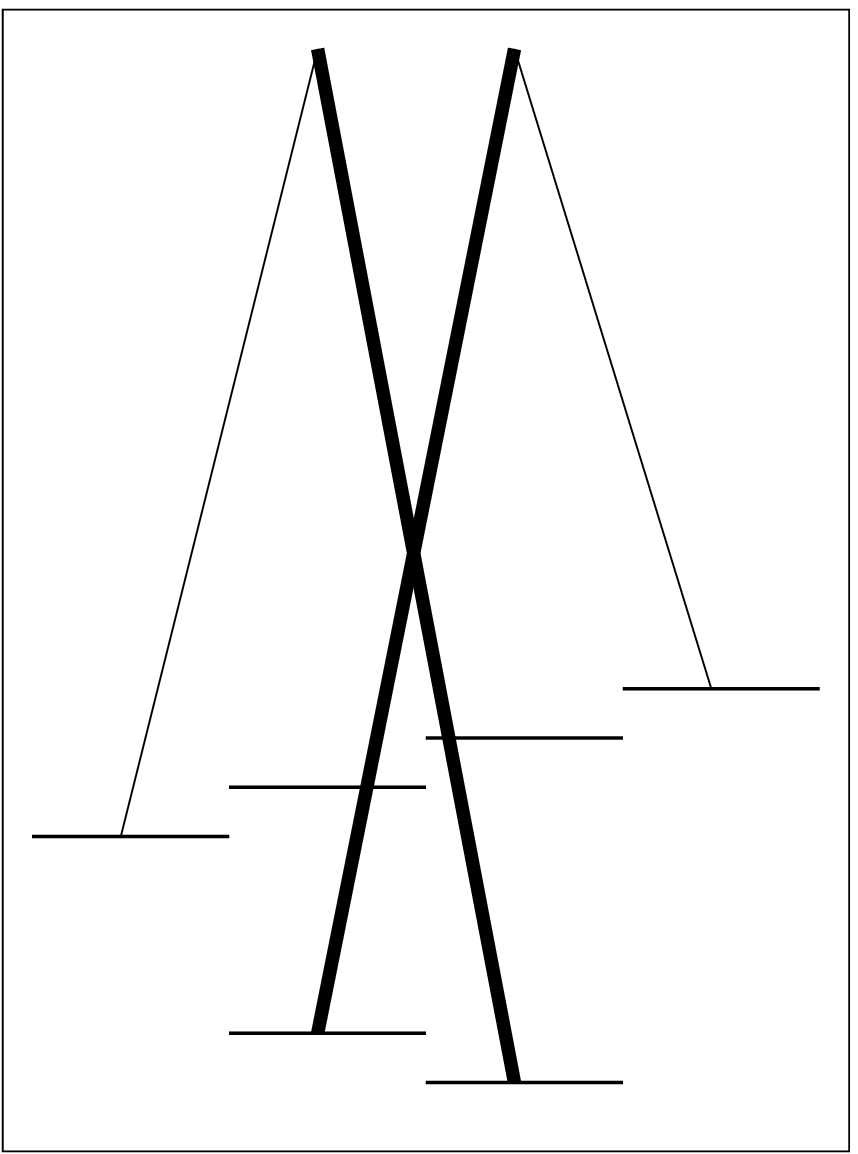}
\end{minipage}
\end{center}
\caption{Fluorescence of the D1 line as a function of frequency
difference. Fixed pump laser tuned to F~=~1/2 $\rightarrow$
F$^{'}$~=~3/2. (a) and (d): B=0, (b) and (e): B~$\approx$~2~G. (a)
and (b): linear perpendicular polarization, (d) and (e): linear
parallel polarization. The transitions shown in (c) and (f) are
responsible for each of the corresponding resonances to their
left. The thicker lines represent the pump transitions.}
\label{fig:EITab}
\begin{picture}(0,0)
\put(22,289){\scriptsize(a)} \put(22,179){\scriptsize(d)}
\put(164,289){\scriptsize(b)} \put(164,179){\scriptsize(e)}
\put(288,287){\scriptsize(c)} \put(288,176){\scriptsize(f)}
\end{picture}
\begin{picture}(0,0)
\put(-10,252){$\Uparrow\rightarrow$}
\put(-10,140){$\Uparrow\uparrow$} \put(63,302){B=0}
\put(205,302){B$\neq$0}
\end{picture}
\end{figure}

Figure~\ref{fig:EITab} shows several EIT resonances obtained for
different polarizations with and without an applied magnetic field
($\approx$~2~G) along the atomic beam. In
figure~\ref{fig:EITab}~(a) and figure~\ref{fig:EITab}~(b) the
polarization of both lasers is linear and perpendicular to each
other. The two outer peaks in figure~\ref{fig:EITab}~(b) are both
shifted by 2$\Delta$ from the zero field resonance which is
consistent with the Raman transitions depicted in
figure~\ref{fig:EITab}~(c). In figure~\ref{fig:EITab}~(d) and
figure~\ref{fig:EITab}~(e) the polarization of both laser beams
are aligned parallel to each other and orthogonal to the magnetic
field. Only two $\Lambda$-type Raman transitions are possible in
this configuration (figure~\ref{fig:EITab}~(f)). The two resulting
EIT resonances are both shifted by $\Delta$ to either side of the
zero-magnetic field resonance. The amplitudes are much smaller
when parallel polarization light was applied. This can be
understood by considering that the atoms will undergo transitions
between all Zeeman sublevels, but for parallel polarized pump and
probe light, not all of these contribute to coherences (see
figure~\ref{fig:EITab}~(f)). Similar results were obtained for the
D2 line.

\begin{figure}[htb]
\begin{center}
\begin{minipage}{0.49\linewidth}
\center
\includegraphics[width=1\linewidth,height=0.8\linewidth]{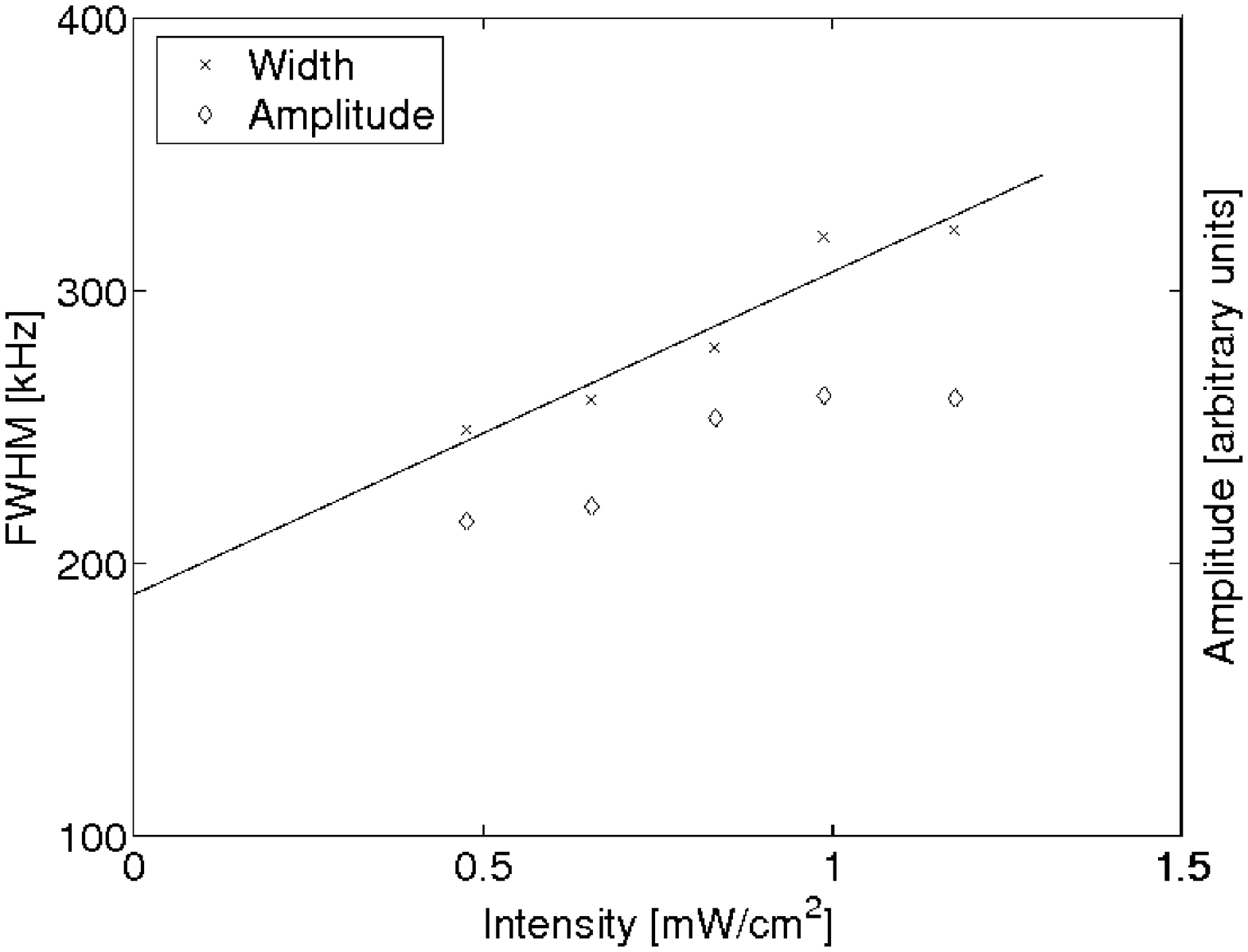}
\end{minipage}
\begin{minipage}{0.49\linewidth}
\includegraphics[width=1\linewidth,height=0.8\linewidth]{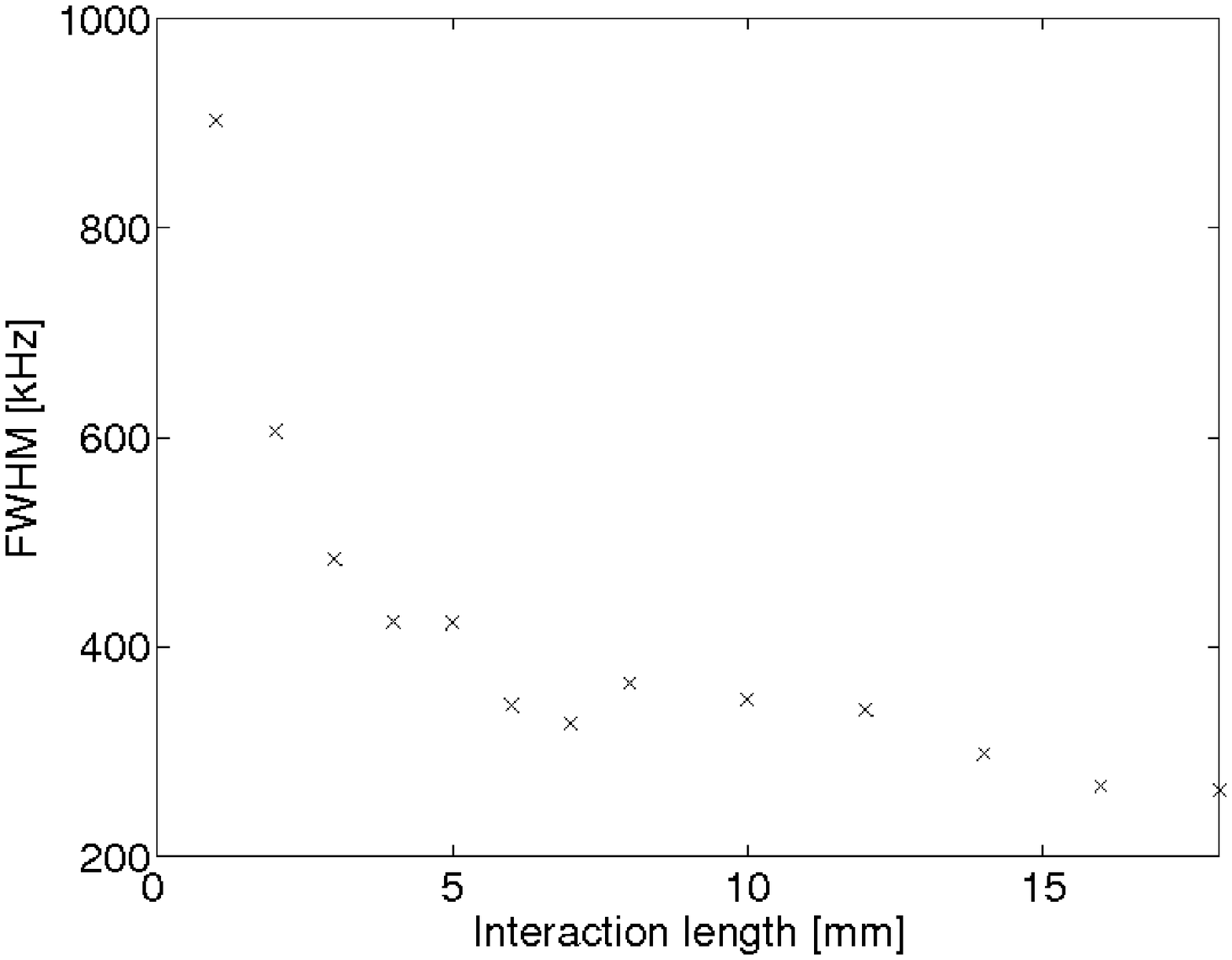}
\end{minipage}
\end{center}

 \caption{(a) Width of observed EIT resonance versus
pump laser intensity. The intensity of the probe laser was fixed
at 1~mW/cm$^2$. The amplitude of the resonances as a function of
laser intensity is also shown. (b) Width of observed EIT resonance
versus interaction length. The sum of the laser intensities was
1.7~mW/cm$^2$. In (a) and (b) EIT was achieved using the
$^2$P$_{1/2}$, F$'$~=~3/2 excited state and the polarization of
the laser beams was linear and perpendicular to each other.}
\label{fig:powbroad}
\begin{picture}(0,0)
\put(155,215){\scriptsize(a)} \put(342,216){\scriptsize(b)}
\end{picture}
\end{figure}

The EIT resonances in the $^{6}$Li beam (figures~\ref{fig:EIT1232}
and~\ref{fig:EITab}) have a width of $\approx$ 300~kHz for the D1
line. This is a factor of 10 reduction compared to the vapour cell
experiments. To get a better understanding of the broadening of
our EIT resonances in the atomic beam we investigated the
dependence of the width on the laser intensity of the pump laser,
as shown in figure~\ref{fig:powbroad}~(a). In addition, the
amplitudes of the EIT resonances are shown. Note that the
intensity of the probe laser is kept fixed at 1~mW/cm$^2$. We
observe a linear dependence of both the amplitude and the
resonance width on the laser intensity. The linear behaviour of
the resonance width is expected when power broadening becomes
significant \cite{Agapev93}. Extrapolation gives a low-intensity
limit of 190~kHz.

Additional broadening comes from the limited interaction time of
the atoms with the light field (transit-time broadening). This was
studied by measuring the resonance width for different laser beam
cross-sections and the results are shown in
figure~\ref{fig:powbroad}(b). In our work the sum of the laser
intensities was kept fixed at 1.7~mW/cm$^2$. A significant
increase could be observed for interaction lengths smaller than
5~mm. However, for the beam diameter of 18~mm used in all other
experiments the broadening due to the transit time is negligible
relative to the contribution of power broadening and magnetic
field inhomogeneities.

\section{Ramsey Spectroscopy}
Further reduction of EIT resonance widths was achieved using
Ramsey spectroscopy \cite{Ramsey89}. In this technique atoms or
molecules in a beam pass two spatially separated interaction
regions. Optical Ramsey fringes were observed on beams of Na and
Cs atoms pumped into coherent non-absorbing states
\cite{Thomas82,Hemmer93}.

Here, a coherent superposition of the $F=1/2$ and $F=3/2$ ground
states is prepared in a light field. The laser field consists of
two co-propagating laser beams which are resonant with the $F=1/2
\rightarrow F'=3/2$ and $F=3/2 \rightarrow F'=3/2$ transitions.
After some time the state is probed in a second interaction region
consisting of the same two frequencies. Similar to the EIT
experiments described previously, the laser resonant with the
$F=1/2 \rightarrow F'=3/2$ transition is kept fixed while the
other frequency is swept over the $F=3/2 \rightarrow F'=3/2$
transition.

\begin{figure}[h]
\begin{center}
\epsfig{file=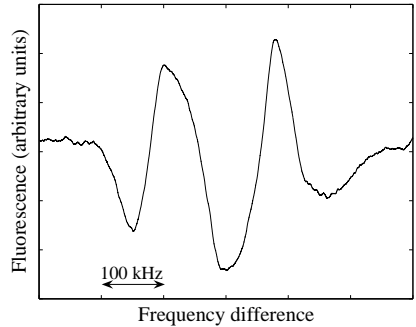,width=0.49\linewidth}
\end{center}
\caption{Optical Ramsey fringes of two co-propagating laser beams
in a Raman-type configuration.} \label{fig:RF}
\begin{picture}(0,0)
\end{picture}
\end{figure}

In our experiment the two interaction regions were separated by
$L$~=~7.7~mm. This was achieved by using only one laser beam in
which the complete central vertical portion was blocked.
Figure~\ref{fig:RF} shows a plot of fluorescence as a function of
frequency difference. The intensities for both pump and probe were
2~mW/cm$^2$. The polarization of both beams were linear and
perpendicular to each other. To amplify the Ramsey fringes the
laser light in the first interaction region was chopped, while the
fluorescence of the probe region was detected by a photo
multiplier tube using a lock-in amplifier. The obtained width
(FWHM) of the resonance was narrower than 100~kHz which is
consistent with calculations based on a mean atomic velocity
$v~\approx$~1600~ms$^{-1}$ and a laser field separation $L$:
$\Delta\nu=v/(3L)=70$~kHz \cite{Demtroeder03}.

\section{Summary}

We have reported electromagnetically induced transparency in
$^{6}$Li in both the vapour cell and atomic beam. Resonant light
with two frequency components was used to produce coherences
between the two hyperfine levels in the $^6$Li ground state for
the first time, allowing fluorescence resonances as narrow as
200~kHz to be observed. The effects of various optical
polarization configurations and applied magnetic fields were
investigated, and these were readily interpreted with
$\Lambda$-type Raman transitions. It has been found that the
maximum contrast of the sub-natural resonances of suppressed
fluorescence occurs for orthogonal linear Zeeman polarizations of
the probe and pump components. Magnetic field insensitive
fluorescence resonance (in the linear approximation) has been
demonstrated despite the fact that the
m~=~0~$\rightarrow$~m$^{'}$~=~0-type Raman transition does not
exist for $^6$Li.

The EIT resonances for the D2 line have been observed in spite of
destructive excitation via cycling transition. However, the
sub-natural resonances are weaker and broader compared to the D1
line. Additional experiments using the Ramsey separated field
technique showed further reduction in the width of the EIT
resonances. As an alkali atom with integer nuclear spin, fermionic
$^6$Li presents a potentially interesting system in which to
investigate coherence effects such as EIT. Its energy level scheme
differs from other atomic species previously investigated, in
terms of the number of hyperfine excited states and their small
energy splitting. The results reported here are consistent with
coherence effects observed using other alkali atoms, such as Rb
and Cs, although the smaller hyperfine splitting of $^6$Li has
allowed significant experimental simplification. Our initial
atomic coherence investigations on this previously unexplored
isotope provide a basis for additional studies.

{\ack} We would like to thank Peter Hannaford for his useful
advice and discussions. This project is supported by the
Australian Research Council Centre of Excellence for Quantum-Atom
Optics and Swinburne University of Technology. \\ \\

\section*{References}
\begin{harvard}

\bibliographystyle{phaip}

\harvarditem{Agap'ev}{1993}{Agapev93} Agap'ev B D, Gornyi M B,
Matisov B G, Rozhdestvenskii Yu V 1993 {\it Usp. Fiz. Nauk} {\bf
163}~1--36

\harvarditem{Akulshin \etal}{2003}{Akulshin03} Akulshin A M,
Cinnnino A, Sidorov A I, McLean R and Hannaford P 2003 {\it
Journal of Optics B-Quantum and Semiclassical Optics} {\bf
5}~S479--S485

\harvarditem{Arimondo}{1996}{Arimondo96} Arimondo E 1996 {\it
Prog. Opt.} {\bf 30}~257--354


\harvarditem{Demtr\"oder}{2003}{Demtroeder03} Demtr\"oder W 2003
{\it Laser spectroscopy: Basic concepts and Instrumentation} vol~3
(New Delhi: Springer-Verlag)

\harvarditem{Harris}{1997}{Harris97} Harris S E 1997 {\it Physics
Today} {\bf 50(7)}~32

\harvarditem{Harris and Hau}{1999}{Harris99} Harris S E and Hau L
1999 {\it Phys. Rev. Lett.} {\bf 82}~4611--4614

\harvarditem{Hemmer \etal}{1993}{Hemmer93} Hemmer P R, Shahriar M
S, Lamela-Rivera H, Smith S P, Bernacki B E and Ezekiel S 1993
{\it J. Opt. Soc. Am. B} {\bf 10}~1326

\harvarditem{Knappe \etal}{2005}{Knappe05} Knappe S, Schwindt P D
D, Shah V, Hollberg L, Kitching J, Liew L and Moreland J 2005 {\it
Optics Express} {\bf 13}~1249--1253

\harvarditem{Lukin}{2003}{Lukin03} Lukin M D 2003 {\it Rev. Mod.
Phys.} {\bf 75}~457--472

\harvarditem{Magnus \etal}{2005}{Magnus05} Magnus F, Boatwright A
L, Flodin A and Shiell R C 2005 {\it J. Opt. B:Quantum Semiclass.
Opt.} {\bf 7}~109--118

\harvarditem{Matsko \etal}{2001}{Matsko01} Matsko A B,
Kocharovskaya O, Rostovtsev Y, Welch G R, Zibrov A S and Scully M
O 2001 {\it Advances in Atomic Mol. and Opt. Phys.} {\bf
46}~191-242

\harvarditem{McAlexander \etal}{1996}{McAlexander96} McAlexander W
I, Abraham E R I, and Hulet R G 1996 {\it Phys. Rev. A} {\bf
54}~R5--R8


\harvarditem{Ramsey}{1989}{Ramsey89} Ramsey N F 1989 {\it
Molecular Beams} 2nd Edn, Clarendon, Oxford

\harvarditem{Schmidt \etal}{1996}{Schmidt96} Schmidt O, Wynands R,
Hussein Z and Meschede D 1996 {\it Phys. Rev. A} {\bf 53}~R27–-R30

\harvarditem{St\"ahler \etal}{2002}{Stahler02} St\"ahler M,
Wynands R, Knappe S, Kitching J, Hollberg L, Taichenachev A and
Yudin V {\it Opt. Lett.} {\bf 27}~1472-1474

\harvarditem{Thomas \etal}{1982}{Thomas82} Thomas J E, Hemmer P R,
Ezekiel S, Leiby C C Jr., Picard R H and Willis C R 1982 {\it
Phys. Rev. Lett.} {\bf 48}~867--870

\harvarditem{Walls \etal}{2003}{Walls03} Walls J, Ashby R, Clarke
J J, Lu B and van Wijngaarden W A 2003 {\it European Physics
Journal D} {\bf 22}~159--162

\harvarditem{Wynands \etal}{1999}{Wynands99} Wynands R and Nagel A
1999 {\it Appl. Phys. B} {\bf 68}~1--25

\harvarditem{Wijngaarden}{2005}{Wijngaarden05} van Wijngaarden W A
2005 \CJP {\bf 83}~327

\harvarditem{Zibrov \etal}{1995}{Zibrov95} Zibrov A S, Lukin M D,
Nibonov D E, Hollberg L, Scully M O, Velichansky V L and Robinson
H G 1995 {\it Phys. Rev. Lett.} {\bf 75}~1499--1502

\end{harvard}


\end{document}